\documentclass[pra,twocolumn,showpacs,preprintnumbers,amsmath,amssymb,superscriptaddress]{revtex4}
\usepackage{revsymb}
\usepackage[cp1250]{inputenc}   %windows
\usepackage[T1]{fontenc}
\usepackage{amssymb,amsmath,amscd,amsfonts}
\usepackage[dvips]{graphicx}% Include figure files
\usepackage{dcolumn}% Align table columns on decimal point
\usepackage[FIGTOPCAP,raggedright,nooneline,bf,footnotesize]{subfigure}
\usepackage{indentfirst}
\usepackage[usenames]{color}
\numberwithin{equation}{section}% [if desired]

\newcommand{\mathsym}[1]{{}}

\newcommand{\V}[1]{\boldsymbol{#1}}
\newcommand{\VV}[1]{\text{\textbf{#1}}}
\newcommand{\kp}[1]{\text{\textbf{k}}_{#1 \perp}}

\renewcommand{\eqref}[1]{Eq.~(\ref{#1})}
\newcommand{\figref}[1]{Fig.~\ref{#1}}
\newcommand{\secref}[1]{Sec.~\ref{#1}}

\newcommand{\eq}[1]{Eq.~(\ref{#1})}

\renewcommand{\vec}[1]{\mathbf{#1}}
\newcommand{\mc}[1]{\mathcal{#1}}

\newcommand{\allarg}{\kp{s},\omega_s;\kp{i},\omega_i}
\newcommand{\pur}{\mathcal{P}}

\newcommand{\thCGA}{\Theta^{(\text{C})}}
\newcommand{\thP}{\Theta^{(\text{P})}}

\newcommand{\phP}{\Phi^{(\text{P})}}
\newcommand{\phD}{\Phi^{(\text{D})}}

\newcommand{\atemp}{A_p^{\text{temp}}}
\newcommand{\asp}{A_p^{\text{sp}}}

\newcommand{\blroz}{\mathbf{d}}

\newcommand{\RcD}{R_c^{(\text{D})}}
\newcommand{\RcP}{R_c^{(\text{P})}}

\newcommand{\tp}{\tau_p^{\text{FWHM}}}

\begin{document}
\pacs{42.65.Lm,42.50.Dv,03.67.Bg}
% 42.65.Lm Parametric down conversion and production of entangled photons
% 42.50.Dv Quantum state engineering and measurements
% 03.67.Bg Entanglement production and manipulation

\title{Modelling and optimization of photon pair sources based on spontaneous parametric down-conversion}
\author{Piotr Kolenderski}
\email{kolenderski@fizyka.umk.pl}
%\homepage{http://www.fizyka.umk.pl/~kolenderski}
\affiliation{Institute of
Physics, Nicolaus Copernicus University, Grudziadzka 5, 87-100
Toru{\'n}, Poland}
\author{Wojciech Wasilewski}
\affiliation{Institute of Experimental Physics, Warsaw University,
Ho{\.z}a 69, 00-681 Warsaw, Poland}
\affiliation{Niels Bohr Institute, University of Copenhagen, DK 2100, Denmark}
\affiliation{QUANTOP, Danish National Research Foundation Center for
Quantum Optics}
\author{Konrad Banaszek}
\affiliation{Institute of Physics, Nicolaus Copernicus University, Grudziadzka 5, 87-100 Toru{\'n}, Poland}
\date{\today}

\begin{abstract}
We address the problem of efficient modelling of photon pairs generated in spontaneous parametric down-conversion and coupled into single-mode fibers. It is shown that when the range of relevant transverse wave vectors is restricted by the pump and fiber modes, the computational complexity can be reduced substantially with the help of the paraxial approximation, while retaining the full spectral characteristics of the source. This approach can serve as a basis for efficient numerical calculations, or can be combined with analytically tractable approximations of the phase matching function. We introduce here a cosine-gaussian approximation of the phase matching function which works
for a broader range of parameters than the gaussian model used previously. The developed modelling tools are used to evaluate characteristics of the photon pair sources such as the pair production rate and the spectral purity quantifying frequency correlations. Strategies to generate spectrally uncorrelated photons, necessary in multiphoton interference experiments, are analyzed with respect to trade-offs between parameters of the source.
\end{abstract}

\maketitle

\section{Introduction}

Spontaneous parametric down conversion (SPDC) is a nonlinear process in which a pump photon interacting with a crystal decays into two daughter photons. The process has been successfully employed to demonstrate fundamental aspects of quantum mechanics such as the violation of Bell's inequalities \cite{CHSH1969,Kwiat1995}, and
utilized in implementations of quantum teleportation \cite{Boschi1998,Marcikic2003,Ursin2004}, quantum cryptography \cite{Gisin2002}, linear optical quantum information processing \cite{Kok2007}, and other quantum-enhanced technologies.

Typically, photon pairs emerging from non-linear media are described by a complicated spatio-temporal wave function that exhibits correlations in multiple degrees of freedom. In contrast, many applications of photon pairs require their preparation in single isolated spatio-temporal modes. Single spatial modes can be selected by coupling photons into single-mode fibers (SMFs) that in effect filter heavily the SPDC light. Furthermore, many protocols rely on interference between photons originating from independent sources \cite{Kaltenbaek2006,Riedmatten2003}. Although spatial modes are well defined by SMFs, the interference visibility may be compromised by undesirable spectral correlations within individual pairs. One way to tailor the spectral degree of freedom is to use narrowband interference filters, which is easy to implement in an experiment, but in consequence reduces the useful photon flux. An alternative approach is to adjust the setup parameters to enforce the source to produce spectrally uncorrelated pairs  \cite{Dragan2004,URen2005,URen2007,Mosley2008}.

These issues bring the question of optimizing the useful fraction of photon pairs produced by SPDC sources. A purely experimental approach would be just to try various alignments of the source. In practice, this strategy would be rather burdensome owing to the large number of controllable parameters of the setup, their time-consuming adjustments, and long data acquisition times. A natural alternative is to resort to numerical modeling. This however presents its own challenges, as including all relevant degrees of freedom is computationally demanding.

In this paper we discuss approximate methods that alleviate the numerical load necessary to model faithfully realistic SPDC sources. Our approach is based on an observation that optical fibers collecting photons define a relatively narrow range of wave vectors that need to be included in calculations. This justifies applying the paraxial approximation, which makes a substantial portion of the problem tractable analytically. The paraxial approximation can be also combined with a simplification of the two-photon wave function to an analytically manageable form leading to closed formulas. We exploit these strategies to analyze the performance of SPDC sources in quantum information applications.

Coupling of down-converted photons into SMFs has been a subject of a number of works, especially in the case of cw pumping. Kurtsiefer {\em et al.} \cite{Kurtsiefer2001} gave a simple argument showing that careful matching of the SPDC output with the fiber modes increases the collection efficiency. Mathematical models for a cw-pumped source has been derived and compared with experimental data in Refs.~\cite{Bovino2003,Castelletto2004,Castelletto2005}. The collinear case has been analyzed theoretically in Ref.~\cite{Andrews2004}. Dragan \cite{Dragan2004} used a gaussian approximation to model fiber-coupled sources.
The counterintuitive scaling of the production rates with the crystal length has been pointed out by Lee {\em et al.} \cite{Lee2005}, and a detailed analysis of quasi-phase matched structures has been presented by Ljunggren and Tengner \cite{ljunggren2005,ljunggren2006}. More recently, Ling {\em et al.} \cite{Ling2008} provided a method to estimate the absolute emission rates for cw pumping. In the present paper, we concentrate on pulse-pumped SPDC sources and optimization of their performance parameters.

Our numerical calculations incorporate the exact form of dispersion relations for the nonlinear medium and use a second-order expansion of phase mismatch in the transverse wave vectors of the SPDC photons. The modeling is based on two strategies. The first approach resorts to numerical means, but with minimized computational effort that will nevertheless deliver highly accurate results in a broad range of parameters. This method has been used in Refs.~\cite{Wasilewski2006,wasilewski2007}  to compare
experimentally measured characteristics of down-conversion sources with theoretical predictions. The second approach will provide expressions
for the biphoton wave function in a closed analytical form
through a further approximation to the phase matching functions. This approach, which we will call
the {\em cosine-gaussian approximation} (CGA) is based on a more accurate analytically integrable model of the phase matching function than the gaussian model studied previously \cite{URen2003,Dragan2004,URen2005}. We compare both the approaches with direct numerical calculations when no paraxial approximation is applied and all integrals are evaluated by numerical means. As an application of the developed tools, we discuss generation of spectrally uncorrelated photons in a type-I $\beta$-barium borate (BBO) crystal. We consider here two strategies to reduce spectral correlation: one method is to adjust carefully the pump pulse and collection modes, while the other one is to restrict the spectrum of the generated photons with the help of interference filters. We compare source brightness that can be achieved using both methods and relate these results to previous discussions \cite{Dragan2004}.

The paper is organized as follows. In Sec.~\ref{section:TwoPhotonInFreeSpace} we present the setup under consideration  and derive the biphoton wave function in free space. Section \ref{section:NumericalApproach} presents basic assumptions about propagation of a pump beam and output photons and the impact of spatial filtering imposed by SMFs. The biphoton wave function within the paraxial approximation is derived. Next in \secref{section:CGA} we present the cosine gaussian approximation and apply it to derive an analytical formula for the wave function of a photon pair coupled into SMFs. The figures of merit are defined in \secref{section:FiguresOfMerit}, and the approximation of perfect phase matching is used to gain some basic intuitions. Next in Sec.~\ref{section:Comparison} we compare the computational effort and applicability of developed methods. Finally, in \secref{section:SpectrallyUncorrelatedPairs} we analyze strategies to reduce spectral correlations  within photon pairs.

\section{Two-photon wave function}
\label{section:TwoPhotonInFreeSpace}

In the non-degenerate down-conversion process, the pump field, described by the positive-frequency part of the electric field $E^{(+)}_p(\VV{r},t)$, interacts with quantized signal and idler fields, whose creation-operator parts will be denoted as $\hat{E}_s^{(-)}(\VV{r},t)$ and $\hat{E}_i^{(-)}(\VV{r},t)$. The interaction hamiltonian has the form of an integral over the volume $V$ of the crystal \cite{Louisell1961}:
\begin{multline}\label{intred}
    \hat{H}_I(t)=
    \frac{\epsilon_0 \chi^{(2)}}{2}\int\limits_{V}^{} d^3\VV{r}\
    E^{(+)}_p(\VV{r},t) \hat{E}_s^{(-)}(\VV{r},t)
    \hat{E}_i^{(-)}(\VV{r},t) \\ + \text{H.c.},
\end{multline}
where $\epsilon_0$ is the vacuum permittivity and $\chi^{(2)}$ denotes the second-order nonlinear susceptibility coefficient, approximated by a constant.
We will assume that the nonlinear interaction is weak enough to neglect pump
depletion and to justify the first order perturbation theory. We will focus here on type-I phase-matching, when both the down-converted photons have the same polarization direction, perpendicular to that of the pump pulse. The case of type-II phase matching can be analyzed analogously.

\begin{figure}[ht]
     \includegraphics[width=\columnwidth]{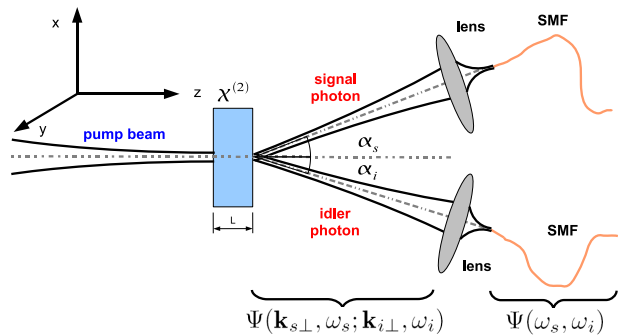}
     \caption{(Color online) The geometry of a photon pair source. A crystal exhibiting $\chi^{(2)}$ nonlinearlity is pumped by gaussian pulse. The generated light emerging at angles $\alpha_s$ and $\alpha_i$ is coupled into single mode optical fibers.} \label{SPDC}
\end{figure}

We will take the nonlinear crystal to be a thin slab
of thickness $L$ oriented perpendicular to $z$-axis and extending from $z=-L/2$ to $z=L/2$,
as illustrated in Fig.~\ref{SPDC}. The pump pulse propagates along $z$--direction
outside the crystal. Following Rubin et al. \cite{klyshko} we
parameterize the waves using the frequencies $\omega$
and the wave vector components $\kp{}$ perpendicular to $z$. These
quantities are preserved at the crystal-free space interface.
In order to calibrate the pump power,
it will be convenient to introduce a normalized pump pulse amplitude $A_p(\kp{p},\omega_p)$ satisfying $\int \text{d}^2\kp{p} \text{d}\omega_p  |A_p(\kp{p}, \omega_p)|^2=1$.
We will assume the pump pulse amplitude in a factorable form, with
no spatiotemporal correlations:
\begin{equation}\label{eq:pumpamplidue}
    A_p(\kp{},\omega)= \atemp(\omega) \asp(\kp{})
\end{equation}
where $\atemp(\omega)$ refers to temporal and  $\asp(\kp{})$ to spatial part. Both parts will be taken in a gaussian form:
\begin{eqnarray}
    \atemp(\omega) &=&   \frac{\sqrt{\tau_p}}{\sqrt[4]{\pi}} \exp\left( -\frac{\tau_p^2}{2 }(\omega-2\omega_0)^2\right)\\
    \asp(\kp{}) &=&   \frac{w_p}{\sqrt{\pi}} \exp\left(-\frac{w_p^2}{2}\kp{}^2 \right)
\end{eqnarray}
where $\tau_p$ stands for the pulse duration, $w_p$ for the pump beam width, and $2\omega_0$ is the central frequency of the pump pulse.

The positive-frequency part of the pump pulse electric field $E^{(+)}_p(\VV{r},t)$ is the Fourier transform of the spectral amplitude:
\begin{equation}\label{pumppulse}
    E^{(+)}_p(\VV{r},t)= \mc{E}_p\int \text{d}^2\kp{p}\text{d} \omega_p A_p (\kp{p},\omega_p) \text{e}^{\text{i}(\kp{p}\VV{r}-\omega_p t )}
\end{equation}
where $\mathcal{E}_p$ characterizes the strength of the pump pulse and the squared modulus $|\mc{E}_p|^2$ proportional to the pump pulse energy.
Subsequently, we assume the following modal expansion for the signal $s$ and idler $i$
field operators:
\begin{multline}\label{fieldop}
    \hat{E}^{(-)}_{\mu}(\VV{r},t) = \mathcal{E}_\mu \int \text{d}^2\kp{\mu}\text{d} \omega_{\mu} \, \text{e}^{-\text{i}\VV{k}_{\mu}\VV{r} + \text{i} \omega_{\mu} t} \hat{a}^\dagger(\kp{\mu},\omega_{\mu}),
    \\ \mu=s,i.
\end{multline}
We approximated here the scaling factors defining the zero-point field fluctuations with frequency-independent constants $\mathcal{E}_\mu$.
The biphoton component of the wave function calculated in the first-order perturbation theory takes the form \cite{Rubin1996}:
\begin{multline}\label{wf:vacume}
    |\Psi\rangle= \frac{1}{\text{i}\hbar}\int \text{d} t\
    \hat{H}_I(t)|\text{vac}\rangle \\
    = \int \text{d}^2\kp{s} \text{d}^2\kp{i} \text{d}\omega_s \text{d}\omega_i \Psi(\allarg) \\
    \times \hat{a}^\dagger\left(\kp{s},\omega_{s}\right)\hat{a}^\dagger\left( \kp{i},\omega_{i}\right)|\text{vac}\rangle
\end{multline}
where the probability amplitude reads:
\begin{multline}\label{eq:wf:intz}
    \Psi(\allarg)= \mc{N} \int_{-L/2}^{L/2}\text{d}z A_p(\kp{s}+\kp{i},\omega_s+\omega_i)
    \\
    \times \exp\left[ \text{i}\Delta k_z(\allarg) z \right]
\end{multline}
and the $\mc{N}={\epsilon_0\chi^{(2)} \mc{E}_p \mc{E}_s
\mc{E}_i }/(2\text{i} \hbar)$. The phase mismatch $\Delta k_z(\allarg)$
is defined using the $z$ components of the wave vectors
of the interacting fields:
\begin{multline}\label{eq:phasemismatching}
    \Delta k_z(\allarg)= \\
    =k_{pz}(\kp{s}+\kp{i},\omega_s+\omega_i) -k_{sz}(\kp{s},\omega_s)-k_{iz}(\kp{i},\omega_i).
\end{multline}
These components are determined by the frequencies $\omega_s,\omega_i$ and the transverse wave vectors $\kp{s},\kp{i}$ \cite{klyshko}. The integral expression in \eq{eq:wf:intz} can be given meaningful
physical interpretation. Each slice of the crystal contributes to a
biphoton amplitude $\Psi(\allarg)$. However,
the phase of this contribution changes from slice to slice, thus only for certain propagation directions the constructive interference occurs.

The wave function given in \eq{eq:wf:intz} describes the entire field emerging from the crystal into the free space. However, in a typical experiment the signal and idler photons are coupled into SMFs. For SMFs collecting light in the $x$--$z$ plane at angles $\alpha_s$ and $\alpha_i$ with respect to the $z$ axis,
the collected spatial modes can be approximated by gaussians centered at transverse wave vectors $\kp{s 0}=\hat x \omega_s \sin \alpha_s /c$ and $\kp{i 0}=-\hat x \omega_i \sin \alpha_i /c$:
\begin{equation}\label{eq:fibermodes}
    u_\mu(\kp{\mu},\omega_\mu)=\frac{w_\mu}{\sqrt{\pi}} \exp\left(-\frac{w_\mu^2}{2}\left(\kp{\mu}-\kp{\mu 0}\right)^2 \right), \quad \mu=s,i
\end{equation}
Here the waists $w_s$ and $w_i$ define the spatial extent of the collected modes, assumed to be constant within the relevant spectral bandwidth.

The wave function $ \Psi(\omega_s,\omega_i)$ for both photons coupled into SMFs is given by an overlap of the wave function in free space $\Psi(\allarg)$ with the spatial profiles $u_s(\kp{s},\omega_s)$ and $u_i(\kp{i},\omega_i)$ of the fiber modes:
\begin{multline}\label{eq:wf:fiber}
    \Psi(\omega_s,\omega_i) = \\
    \int \text{d}^2\kp{s} \text{d}^2\kp{i}\, u_s^*(\kp{s},\omega_s) u_i^*(\kp{i},\omega_i) \Psi(\allarg).
\end{multline}
This object will be used to calculate coincidence count rates and spectral properties of generated photons. For a pump pulse amplitude in a factorable form as that in Eq.~(\ref{eq:pumpamplidue}), it will be convenient to write
\begin{equation}\label{eq:def:theta}
\Psi(\omega_s,\omega_i) = \atemp(\omega_s+\omega_i) \Theta(\omega_s,\omega_i).
\end{equation}
Here $\Theta(\omega_s,\omega_i)$ can be viewed as the effective phase matching function for
the collected modes that includes the geometry of the setup and the physical properties of the nonlinear medium. It is explicitly given by:
\begin{multline}\label{eq:epmf:DNI}
    \Theta(\omega_s,\omega_i) = \mathcal{N}\int \text{d}^2\kp{s}\ \text{d}^2\kp{i}
    \int_{-L/2}^{L/2}\text{d}z \, \asp(\kp{s}+\kp{i}) \\
    \times u_s^*(\kp{s},\omega_s)\ u_i^*(\kp{i},\omega_i)
    \text{e}^{\text{i}
    \Delta k_z(\kp{s},\omega_s;\kp{i},\omega_i) z  }.
\end{multline}
One way to simplify the above equation is to evaluate analytically
the integral over length of the crystal, which yields:
\begin{multline} \label{Eq:empf:D}
    \Theta^{(\text{D})}(\omega_s,\omega_i) =  \frac{\mc{N} L }{2}
    \int \text{d}^2\kp{s}\ \text{d}^2\kp{i}  u_s^*(\kp{s},\omega_s)\ u_i^*(\kp{i},\omega_i) \\
    \times \asp(\kp{s}+\kp{i}) \text{sinc}\left(\frac{L}{2} \Delta k_z(\allarg) \right).
\end{multline}
However, the remaining integrals over $\kp{s}$ and $\kp{i}$ are intractable analytically due to nontrivial form of phase mismatch $\Delta k_z$ and they must be performed by numerical means. We will refer to this procedure as \emph{direct numerical integration} and denote corresponding formulas with a superscript $(D)$. The four-dimensional integration task is computationally very demanding, and in the next two paragraphs we will present approximate methods which reduce the computational effort to compute effective phase matching function $\Theta(\omega_s,\omega_i)$.

\section{Paraxial approximation}
\label{section:NumericalApproach}

The expression for the effective phase matching function given in Eq.~(\ref{eq:epmf:DNI}) includes gaussian fiber mode functions $u_s(\kp{s},\omega_s)$ and $u_i(\kp{i},\omega_i)$ that vanish very fast as the transverse wave vectors $\kp{s}$ and $\kp{i}$ depart from the central observation directions $\VV{k}_{s0\perp}$ and $\VV{k}_{i0\perp}$. This implies that little error is introduced when expanding the phase mismatch $\Delta k_z$ given in \eq{eq:phasemismatching} up to the second order in deviations of the transverse wave vectors from $\kp{s0}$ and $\kp{i0}$. After such an expansion the entire integrand in Eq.~(\ref{eq:epmf:DNI}) takes a gaussian form
in $\kp{s}$ and $\kp{i}$, provided that the spatial pump profile is gaussian as well. Consequently, one can perform all the integrals over transverse wave vectors analytically. This is a great simplification of the computational complexity of the problem, as we are now left only with a one-dimensional integral over $z$ which needs to be performed numerically. We will call this method {\em paraxial approximation} in analogy to the standard description of paraxial wave propagation in classical optics.

It will be convenient to introduce the following notation for the expansion of the wave vector mismatch:
\begin{multline}\label{eq:approx}
    \Delta k_z(\kp{s},\omega_s;\kp{i},\omega_i) \approx \\ \vec{D}_{0}(\omega_s,\omega_i)+\mathbf{D}_{1}^T(\omega_s,\omega_i)\boldsymbol{\kappa} +{\boldsymbol{\kappa}}^T \mathbf{D}_{2}(\omega_s,\omega_i)\boldsymbol{\kappa},
\end{multline}
where
\begin{equation}\label{eq:def:kappa}
    \boldsymbol{\kappa}=(\kp{s}-\kp{s0},\kp{i}-\kp{i0})^T
\end{equation}
is a four-element vector of deviations from the central observation directions. The Taylor expansion coefficients can be grouped into
a scalar in the zeroth order
\begin{equation}
\vec{D}_{0}(\omega_s,\omega_i) = \Delta k_z(\kp{s0},\omega_{s};\kp{i0},\omega_{i})
\end{equation}
a vector in the first order
\begin{equation}
\label{Eq:D1}
\vec{D}_{1}(\omega_s,\omega_i) = \left(\begin{array}{c} \blroz_s(\omega_s,\omega_i)\\ \blroz_i(\omega_s,\omega_i)
\end{array}
\right)
\end{equation}
and a matrix in the second order:
\begin{equation}
\vec{D}_{2}(\omega_s,\omega_i) =
\left( \begin{array}{cc} \blroz_{ss}(\omega_s,\omega_i), & \blroz_{si}(\omega_s,\omega_i) \\ \blroz_{si}(\omega_s,\omega_i), & \blroz_{ii}(\omega_s,\omega_i) \end{array}\right)
\end{equation}
We wrote the vector $\vec{D}_{1}$ and the matrix $\vec{D}_{2}$ in a block form with entries given by:
\begin{equation}
    \blroz_{\mu}(\omega_s,\omega_i) = \left.\left(\frac{\partial \Delta k_z}{\partial k_{\mu x}},\frac{\partial \Delta k_z}{\partial k_{\mu y}}\right)^T\right|_{\scriptsize\begin{array}{c}\kp{s}=\kp{s0}\\ \kp{i}=\kp{i0}\end{array}},
    \end{equation}
and
\begin{multline}
    \blroz_{\mu\nu}(\omega_s,\omega_i) = \\
     =\frac{1}{2}\left.\left(
    \begin{array}{cc} \displaystyle
    \frac{\partial^2 \Delta k_z}{\partial k_{\mu x} \partial k_{\nu x}}, & \displaystyle \frac{\partial^2 \Delta k_z}{\partial k_{\mu x} \partial k_{\nu y}} \\    \displaystyle \frac{\partial^2 \Delta k_z}{\partial k_{\mu y} \partial k_{\nu x}}, & \displaystyle
    \frac{\partial^2 \Delta k_z}{\partial k_{\mu y} \partial k_{\nu y}} \end{array}\right)\right|_{\scriptsize\begin{array}{c}\kp{s}=\kp{s0}\\ \kp{i}=\kp{i0}\end{array}},
\end{multline}
where $\mu, \nu = s,i$.

In order to write a compact formula for the effective phase matching function in the paraxial approximation, it will be helpful to represent the product of the fiber mode functions $u_s^\ast (\kp{s},\omega_s) u_i^\ast(\kp{i},\omega_i)$ and the pump beam profile $\asp(\kp{s}+\kp{i})$ as an exponent of a quadratic expression:
\begin{multline}
    u_s^\ast (\kp{s},\omega_s) u_i^\ast(\kp{i},\omega_i) \asp(\kp{s}+\kp{i})=
    \\
        =\exp\left(-\vec{B}_0-\vec{B}_1^T\boldsymbol \kappa -\boldsymbol \kappa^T \vec{B}_2 \boldsymbol \kappa\right).
\end{multline}
where $ \boldsymbol \kappa$ is a four-element vector of deviations from central observation directions defined in \eqref{eq:def:kappa}. The coefficients of the quadratic expression are a scalar
\begin{equation}
    \vec{B}_0
    =
    \frac{w_p^2}{2}\left(\vec{k}_{s0\perp}+\vec{k}_{i0\perp}\right)^2
\end{equation}
a four-component vector
\begin{equation}
   \vec{B}_1
   =w_p^2 \left(
                    \begin{array}{c}
                    \vec{k}_{s0\perp}+\vec{k}_{i0\perp} \\
                    \vec{k}_{s0\perp}+\vec{k}_{i0\perp}
                    \end{array}
                  \right)
\end{equation}
and a $4\times 4$ matrix
\begin{equation}
     \vec{B}_2=
   \frac{1}{2} \left(
                    \begin{array}{cc}
                    \displaystyle (w_p^2+w_s^2)\vec{I} & \displaystyle w_p^2\vec{I}\\
                    \displaystyle w_p^2\vec{I}  &\displaystyle  (w_p^2+w_i^2)\vec{I}
                    \end{array}
                  \right),
\end{equation}
where $\vec{I}$ denotes a two dimensional identity matrix. This notation allows us to write the result of four-dimensional gaussian integration of Eq.~(\ref{eq:epmf:DNI}) over the transverse wave vectors as:
\begin{multline}\label{eq:epmf:N}
    \thP(\omega_s,\omega_i)= %\atemp(\omega_s+\omega_i)
    \int_{-L/2}^{L/2}\text{d}z\, \frac{\mc{N} w_s w_i w_p}{\sqrt{\pi\det\vec{M}_2(z)}} \\
    \times
   {\exp\left(-\vec{M}_0(z) -\frac{1}{4} \vec{M}^T_1(z) \vec{M}^{-1}_2(z) \vec{M}_1(z)\right)}
\end{multline}
where the superscript $(P)$ stands for the paraxial approximation, we introduced
\begin{equation}
  \vec{M}_j(z)=\vec{B}_j-\text{i}z\vec{D}_{j}, \qquad j=0,1,2
\end{equation}
and for notational simplicity we suppressed dependence on frequencies $\omega_s$ and $\omega_i$. The integral
over the crystal length in Eq.~(\ref{eq:epmf:N}) needs to be calculated numerically, which is substantially faster than direct numerical integration of Eq.~(\ref{Eq:empf:D}).
It is worthwhile to note that in Eq.~(\ref{eq:epmf:N}) the effects of spectral dispersion are fully taken into account, as no expansion in the signal and idler frequencies $\omega_s$ and $\omega_i$ has been applied. As we will see in \secref{section:Comparison}, this makes numerical results based on the paraxial approximation very precise.

\section{Cosine-Gaussian approximation}
\label{section:CGA}

\begin{figure}[h]
    \includegraphics[width=1\columnwidth]{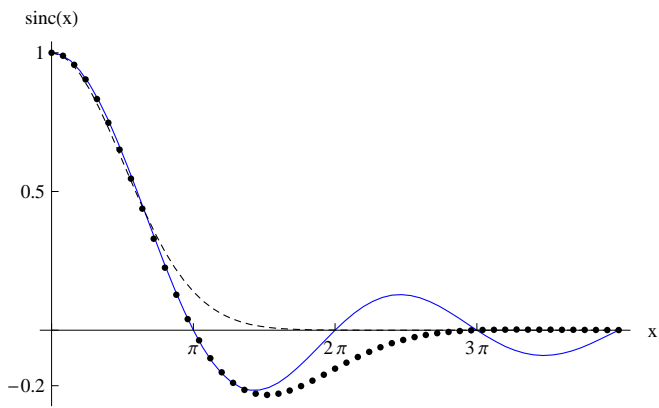}\\
    \caption{(Color online) A comparison of the $\mathop{\text{sinc}}x$ function (solid
    blue line) with the cosine-gaussian approximation $\mathop{\text{sinc}}x \approx \exp{(-x^2/20)\cos (x/2)}$ (circles) and the gaussian approximation
    $\mathop{\text{sinc}}x\approx \exp{(-x^2/5)}$ (dashed line). }\label{fig:sinc}
\end{figure}

The numerical effort to calculate the effective phase matching function can be reduced further at the cost of the accuracy. The basic idea is to replace the sinc term appearing in Eq.~(\ref{Eq:empf:D}) by an analytically tractable expression. Previous works \cite{Dragan2004,URen2005} introduced the {\em gaussian approximation} (GA), which approximated the sinc term by a gaussian function, thus enabling analytical integration. We will consider here a more general expression of the form:
\begin{equation}\label{eq:sincapprox}
    \mathop{\text{sinc}} x\approx \exp(- \xi{x^2} ) \cos
    \left(\zeta x \right) = {\textstyle\frac{1}{2}} \exp(- \xi{x^2} + \text{i} \zeta x) + \text{c.c.}
\end{equation}
As seen in \figref{fig:sinc}, using the parameters $\xi=\frac{1}{20}$ and $\zeta=\frac{1}{2}$ yields a more accurate approximation to the sinc function than the GA corresponding to the choice of parameters $\xi=\frac{1}{5}$ and $\zeta=0$.

The above observation leads us to the idea of {\em cosine-gaussian approximation} (CGA). Specifically, in \eq{Eq:empf:D} we replace the sinc function with \eqref{eq:sincapprox} and expand the phase mismatch $\Delta k_z$ up to the linear term in transverse wave vectors around central observation directions:
\begin{multline}
  \text{sinc}\left(\frac{L}{2} \Delta k_z (\allarg)\right) \approx \\
  {\textstyle\frac{1}{2}}\exp\left(-{\textstyle\frac{1}{4}}\xi(\vec{D}_0+\vec{D}_1^T \V{\kappa})^2 L^2+
   \frac{\text{i}}{2}\zeta (\vec{D}_0+\vec{D}_1^T \V{\kappa})L\right)+\text{c.c.}
\end{multline}
We used here parametrization introduced in Eqs.~(\ref{eq:def:kappa})-(\ref{Eq:D1}).
After inserting the above expression into \eqref{Eq:empf:D}, the integrals over transverse wave vectors can be evaluated analytically as long as the pump and fiber modes are gaussian. This yields an expression for the effective phase matching function of the form:
\begin{multline}\label{eq:epmf:CGA}
    \thCGA(\omega_s,\omega_i)=
    \Gamma(\omega_s,\omega_i) e^{-f(\omega_s,\omega_i)} \cos[g(\omega_s,\omega_i)].
\end{multline}
The three functions appearing in the above formula
are given by:
\begin{eqnarray}
    \Gamma(\omega_s,\omega_i)&=& \frac{\pi^2 \mathcal{N}}{\sqrt{\det \vec{K}}} \\
    g(\omega_s,\omega_i) &=& \frac{1}{2}\zeta  L \vec{D}_0+\frac{1}{16}\zeta  L \vec{D}_1^T \V{K}^{-1}\left(\vec{B}_1+\frac{L^2}{2}\xi \vec{D}_0 \vec{D}_1\right)\\
    f(\omega_s,\omega_i)&=& \vec{B}_0 + \frac{1}{4} \xi L^2 \vec{D}_0^2+\frac{1}{16}\vec{N}^T \vec{K}^{-1} \vec{N}
\end{eqnarray}
where we defined:
\begin{eqnarray}
  \vec{K} &=& \vec{B}_2+\frac{1}{4}\xi L^2\vec{D}_1 \vec{D}_1^T\\
  \vec{N} &=& \vec{B}_1+\frac{L}{2}\left(\xi L \vec{D}_0 +\zeta \right) \vec{D}_1.
\end{eqnarray}

For the sake of brevity we have omitted the frequency dependence. The expression for the effective
 phase matching function in the gaussian approximation is easily obtained by inserting
$\xi= \frac{1}{5}$ and $\zeta=0$.

In order to analyze the applicability of CGA, it is convenient to view the biphoton wave function given in \eq{eq:wf:fiber} as an integral over $\kp{s}$ and $\kp{i}$ of a product of two factors. The first one is the phase matching term $\text{sinc}[\Delta k_z(\allarg) L /2]$, while the second one, which we will call here the beam term, is a triple product of the pump pulse spatio-temporal profile $A_p(\kp{s}+\kp{i},\omega_s+\omega_i)$ and the fiber mode profiles $u_s(\kp{s},\omega_s)$ and $u_i(\kp{i},\omega_i)$. The beam term defines the range of transverse wave vectors and frequencies for which the cosine-gaussian approximation of the phase matching term should be accurate. This condition is satisfied when the sinc argument $\Delta k_z(\allarg) L /2$ does not exceed approximately $3\pi/2$.

Let us analyze this condition more closely. For the profiles assumed throughout this paper, the beam term takes a gaussian form:
\begin{multline}
  A_p(\kp{s}+\kp{i},\omega_s+\omega_i)u_s(\kp{s},\omega_s)u_i(\kp{i},\omega_i) \propto \\
  \exp\left( - \frac{w_p^2 \alpha^2}{2 c^2}(\nu_s-\nu_i)^2- \frac{\tau_p^2}{2}(\nu_s+\nu_i)^2
  -\boldsymbol{\kappa}^T \vec{B}_2 \boldsymbol{\kappa}\right),
\end{multline}
where $\nu_\mu = \omega_\mu - \omega_0$ are detunings from the central frequency and we assumed
that the photons are collected at identical angles $\alpha_s=\alpha_i=\alpha$. In the exponent, we neglected the cross-term correlating wave vectors with frequencies.

The characteristic width of the Gaussian function defines the relevant range of parameters. Thus the sum of the detunings is restricted by $|\nu_s+\nu_i| \lesssim  \tau_p^{-1}$. Similarly the range of relevant transverse wave vectors can be crudely characterized by the smallest eigenvalue of the matrix $\vec{B}_2$, which is equal to $w_s^2$ in case of symmetric coupling $w_s=w_i$. This can be written as $|\boldsymbol{\kappa}|\lesssim w_s$. In the case of perfect phase matching for the central wave vectors $\kp{s0}$, $\kp{i0}$ at the frequency $\omega_0$ of the down-converted photons, we estimate the argument of the sinc function expanding
the wave vector mismatch $\Delta k_z$ up to the first order:
\begin{equation}\label{eq:CGA:phasemismatch}
     \Delta k_z \approx \vec{D}_1(\omega_0,\omega_0)\V{\kappa}+ \beta(\nu_s+\nu_i) ,
\end{equation}
where $\beta= \left.\frac{\partial k_{pz}}{\partial \omega_p}\right|_{\omega_p=2\omega_0}- \left.\frac{\partial k_{sz}}{\partial \omega}\right|_{\omega=\omega_0} $. Thus we see that the CGA will be valid, if
$|\V{\kappa}|$ and $|\nu_s+\nu_i|$ within ranges defined by the beam term yield the argument of the sinc function $\lesssim 3\pi/2$. This gives:
\begin{equation}
\tau_p \gtrsim \beta L
\end{equation}
and
\begin{equation}
w_s \gtrsim  L |\vec{D}_1 |
\end{equation}
As the right hand sides in the above formulas are estimates, we rounded up numerical factors to simpler forms.

\section{Figures of merit}\label{section:FiguresOfMerit}

We will employ the computational methods presented in the preceding sections to analyze two parameters characterizing the usefulness of SPDC sources. The first one is the brightness, proportional to the probability of producing a
fiber-coupled photon pair by a single pump pulse:
\begin{equation}\label{eq:def:rc}
    R_c = \int \text{d}\omega_s  \text{d}\omega_i\, | \Psi(\omega_s,\omega_i)|^2.
\end{equation}
We will set the brightness unit by putting the multiplicative factor appearing in
\eq{eq:wf:intz} to be $|\mc{N}|=1$.

The second important property of photon pairs is their suitability for multiphoton interference experiments. When interfering photons from independent sources, their spectral amplitudes cannot carry any distinguishing information about the origin of the photons. This means that the biphoton wave function for each pair should be factorable. The degree of factorability can be quantified with the help of the
Schmidt decomposition, which for the normalized wave function
$\Psi(\omega_s,\omega_i)/\sqrt{R_c}$ takes the form \cite{law2000}:
\begin{equation}\label{eq:schmodt}
    \frac{1}{\sqrt{R_c}} \Psi(\omega_s,\omega_i)=\sum_{n=0}^\infty
    \sqrt{\varsigma_n} \phi_n^s(\omega_s)\phi_n^i(\omega_i).
\end{equation}
In the above expression, $\phi_n^s(\omega_s)$ and $\phi_n^i(\omega_i)$ are two orthonormal sets
of mode functions for the signal and the idler photons. The nonnegative parameters $\varsigma_n$ characterize the contribution of each pair of modes to the superposition. They satisfy the normalization constraint $\sum_{n=0}^{\infty}\varsigma_n = 1$ and it is convenient to put them in the decreasing order. Perfect factorability thus corresponds to the condition $\varsigma_0 =1$.

The degree of factorability can be quantified by the visibility of two-photon interference. Suppose that two heralded signal photons produced by identical sources are superposed on a 50:50 beamsplitter and the depth of the Hong-Ou-Mandel dip \cite{Hong1987} is measured. The depth is given by a nonnegative expression
\begin{equation}\label{eq:def:purity}
  \pur=\sum_{n=0}^\infty \varsigma_n^2.
\end{equation}
which will be called the \emph{purity parameter} of a photon pair.
In general $\pur\le 1$ and the equality sign holds only for a factorable biphoton wave function. The purity parameter is the inverse of cooperativity parameter introduced in Ref.~\cite{Huang1993}.

Typically, photon pairs are spectrally filtered in order to improve their characteristics and
to lower the background count rates. The effects of  spectral filtering can be taken into account by multiplying the two-photon wave function by spectral amplitude transmissions $\Lambda_\mu(\omega_\mu)$ characterizing the filters:
\begin{equation}\label{eq:subst:bi}
  \Psi(\omega_s,\omega_i)\rightarrow {\Lambda_s(\omega_s)\Lambda_i(\omega_i)}\Psi(\omega_s,\omega_i)
\end{equation}
Note that the above substitution correctly takes into account the decrease in count rates resulting from spectral filtering. We will model spectral filters using gaussian profiles with respective widths $\sigma_s$ and
$\sigma_i$, assuming perfect transmission at the peak frequency $\omega_0$:
\begin{equation}\label{eq:subst:one}
    \Lambda_\mu(\omega)=\exp\left(-\frac{(\omega-\omega_0)^2}{ 2 \sigma_\mu^2 }\right),\quad \mu=s,i
\end{equation}
It is worthwhile to stress that the spatial filtering imposed by SMFs and spectral filtering implemented with interference filters are of different nature. The SMFs perform coherent filtering at the field level, i.e.\ add field amplitudes, while spectral filters transmit independently each frequency component.

Before discussing characteristics of realistic sources, it is insightful to consider the
limit of perfect phase matching, based on an assumption that
$\Delta k_z(\allarg) L/2 \approx 0$ over the relevant range of frequencies and wave vectors. This approximation means that we can put $\vec{D}_0= \vec{D}_1=\vec{D}_2=0$, which makes the integrand in \eqref{eq:epmf:N} independent of $z$ and leads to a very simple formula for the fiber-coupled biphoton wave function:
\begin{multline}\label{eq:wf:zero}
    \Psi^{(0)}(\omega_s,\omega_i)= \\
    4\mathcal{N}\sqrt[4]{\pi}\frac{ L   \bar w^2 \sqrt{\tau_p}} {w_i w_p w_s}
    \exp\left(-\frac{ n_o^2(\omega_0) \bar{w}^2 }{2c^2}  (\omega_s\alpha_s-\omega_i\alpha_i)^2 \right)
    \\
    \times \exp \left(-\frac{\tau_p^2}{2}(\omega_s +\omega_i
    -2\omega_0)^2\right)
\end{multline}
where by the superscript $(0)$ we indicated the approximation of perfect phase matching. We also took the refractive indices at the central frequency $n_o(\omega_s)\approx n_o(\omega_i)\approx n_o(\omega_0)$ and denoted
\begin{equation}
  \bar{w}=\left(\frac{1}{w_s^2}+\frac{1}{w_i^2}+\frac{1}{w_p^2} \right)^{-1/2}.
\end{equation}
 Let us note that the assumption $\Delta k_z L/2 \approx 0$ implies a specific geometry of the setup. First, it means that the pump, signal and idler beams maintain good spatial overlap through the entire length of the crystal. Secondly, the length $L$ of the crystal must be much shorter than the characteristic Rayleigh range of the beams.

The wave function given in \eqref{eq:wf:zero} is gaussian, which leads to closed analytical formulas for parameters of interest.
The brightness can be easily calculated to be equal to:
\begin{equation}\label{eq:zero:rc}
    R_c^{(0)}= \frac{16 \pi^{3/2} c L^2}{n_0(\omega_0)(\alpha_s+\alpha_i)}\frac{\bar w^3}{ w_s^2 w_i^2 w_p^2}
\end{equation}
It is instructive to analyze the scaling of the pair production rate in the parameters involved. The quadratic dependence on the crystal length $L$ is a result of a coherent summation of the probability amplitudes of generating a photon pair over the entire range of $-L/2 \le z \le L/2$.
Assuming that the waists of the pump, signal, and idler beams are of the same order characterized by $w$, the pair production rate scales as $1/w^3$. This scaling can be interpreted as a result of an interplay of two effects. The first one is the dependence of the nonlinear process on the transverse spatial dimension of the interacting modes. Suppose that the modes are confined to a transverse area of the order of $w^2$. Then their normalization includes a factor $1/w$ for each of the modes. As the probability amplitude for pair generation involves an integral of a product of three mode functions over an area of size $w^2$, this gives its scaling as $1/w$. Squaring this result gives the probability of pair generation scaling as $1/w^2$. The second effect is the broadening of the spectrum of the produced photons with decreasing waists seen in the first exponent in \eqref{eq:wf:zero}, which yields an additional factor of $1/w$.

The expression calculated in \eqref{eq:zero:rc} enables us to optimize the pair production rate with respect to some parameters of the setup. For example, suppose that the waists $w_s$ and $w_i$ of the collection modes are fixed. An easy calculation shows that the maximum production rate is achieved for the pump beam waist $w_p$ given by:
\begin{equation}\label{eq:zero:optimalwaist}
    w_p=\frac{w_s w_i}{\sqrt{2(w_s^2 +w_i^2)}}
\end{equation}
which reduces to $w_p=w_s/2$ for equal waists of collection modes. We will use this coupling strategy through the rest of the article. Note that in the case of a monochromatic pump, in crude approximation of perfect phase mismatch the condition for optimal brightness for short crystal lengths takes the form $w_p=w_s/\sqrt{2}$ \cite{Ling2008}.

As noted in Refs.~\cite{URen2003, URen2005}, in the approximation of perfect phase matching the condition
for spectral decorrelation within a photon pair is achieved when
\begin{equation}\label{eq:zero:dec}
  \tau_p=\frac{w_p\, \alpha_s \alpha_i}{c}.
\end{equation}
A more general analytical condition can be derived using the gaussian approximation \cite{Dragan2004}. Within this model the biphoton wave function takes following form:
\begin{equation}\label{eq:c1}
    \Psi^{(\text{G})}(\omega_s,\omega_i)=\sqrt{\frac{\tau_p}{\sqrt{\pi}}}\Gamma(\omega_s,\omega_i) e^{-f(\omega_s,\omega_i)-{\tau_p^2}(\omega_s +\omega_i -2\omega_0)^2/2}
\end{equation}
Taking $\Gamma(\omega_s,\omega_i)\approx\Gamma(\omega_0,\omega_0)$ and expanding $f(\omega_s,\omega_i)$ up to the second order in frequencies around $\omega_0$ yields a gaussian expression in detunings. Spectral decorrelation corresponds to the vanishing cross-term $(\omega_s-\omega_0)( \omega_i-\omega_0)$ in the exponent, which gives:
\begin{equation}\label{eq:tpdec}
  \tau_p^2=\left.2\frac{ \partial^2 f (\omega_s,\omega_i)}{\partial\omega_s\partial\omega_i}\right|_{\omega_s=\omega_i=\omega_0}.
\end{equation}
More accurate models of the effective phase matching function in Eqs.~(\ref{eq:epmf:N}) and (\ref{eq:epmf:CGA})
do not yield a decorrelation condition in a closed analytical form.

\section{Comparison}\label{section:Comparison}

\begin{figure}
\begin{tabular}{cc}
  \subfigure[$L=100~\mu$m]{\includegraphics[width=0.49\columnwidth]{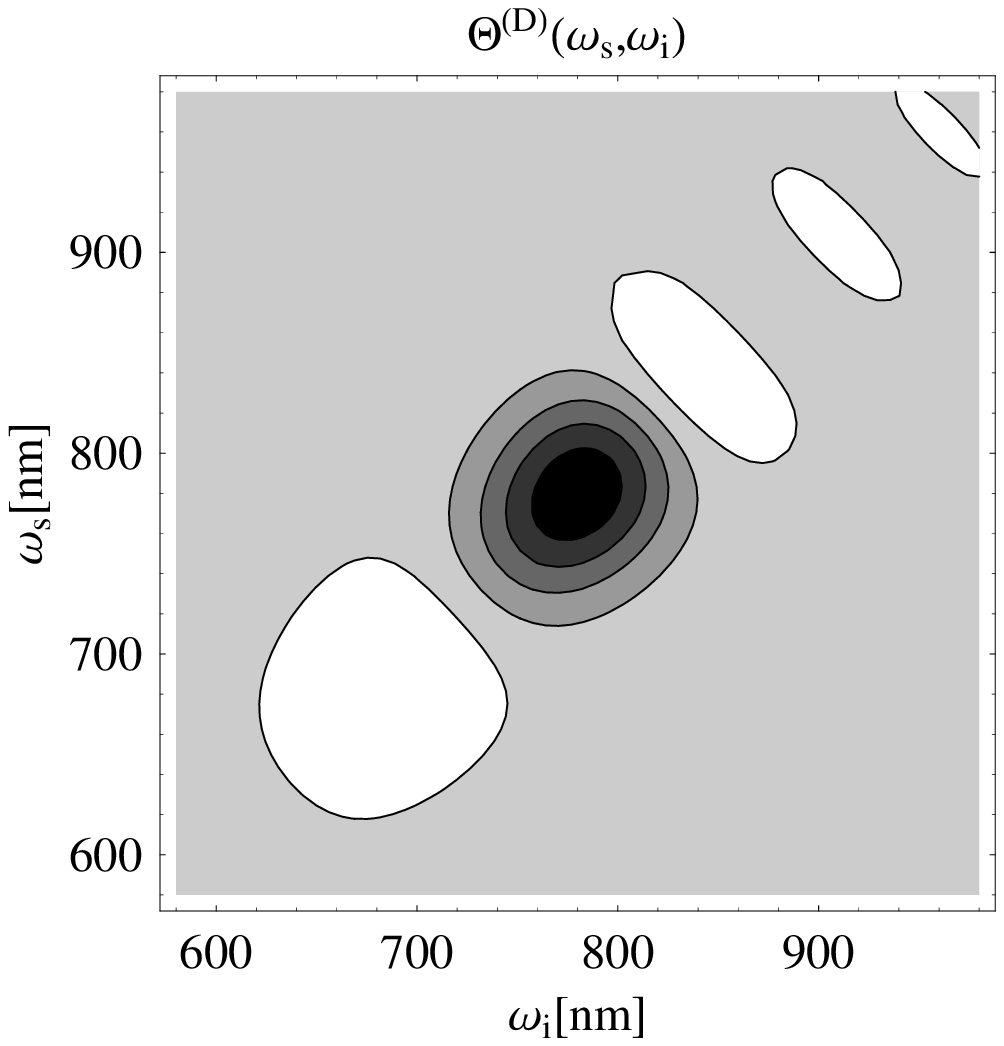}}&
  \subfigure[$L=1$~mm]{\includegraphics[width=0.49\columnwidth]{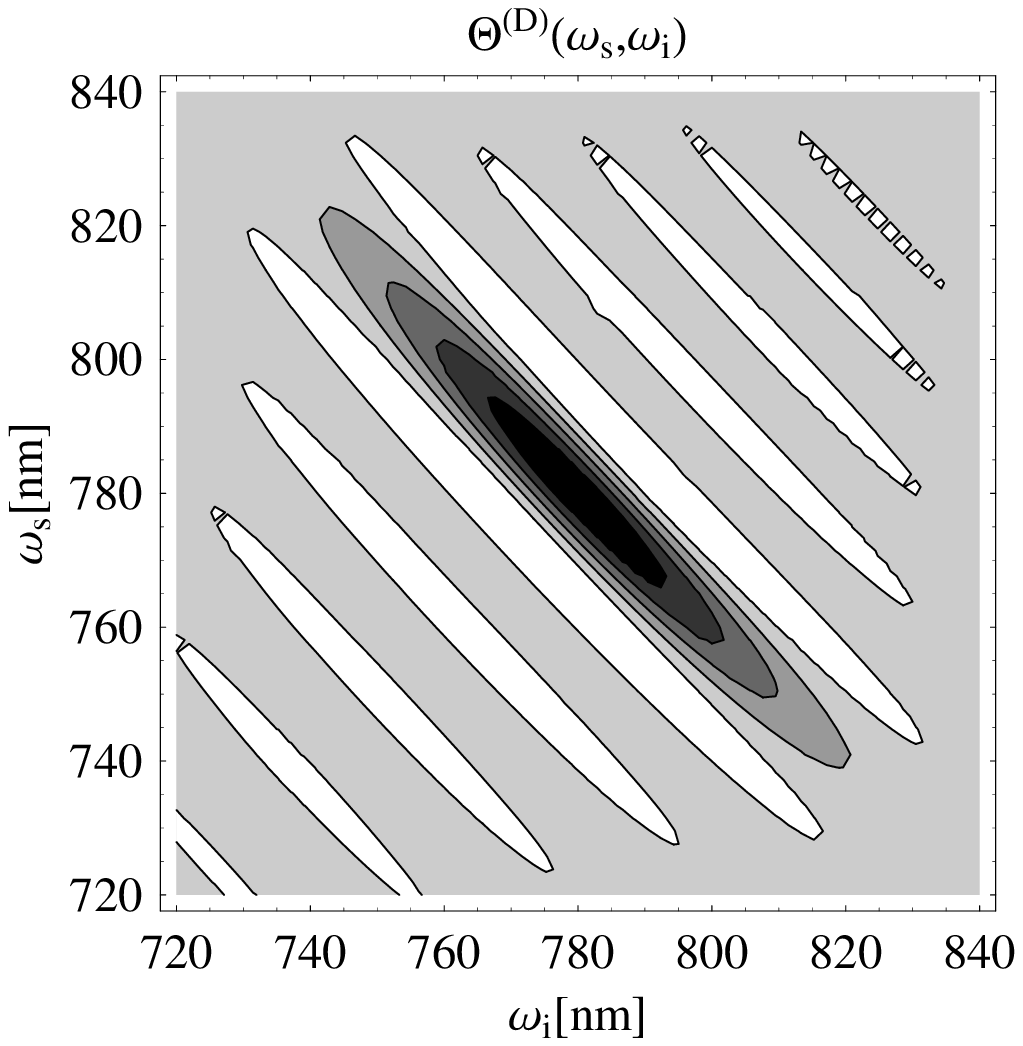}}\\
  \subfigure[]{\includegraphics[width=0.49\columnwidth]{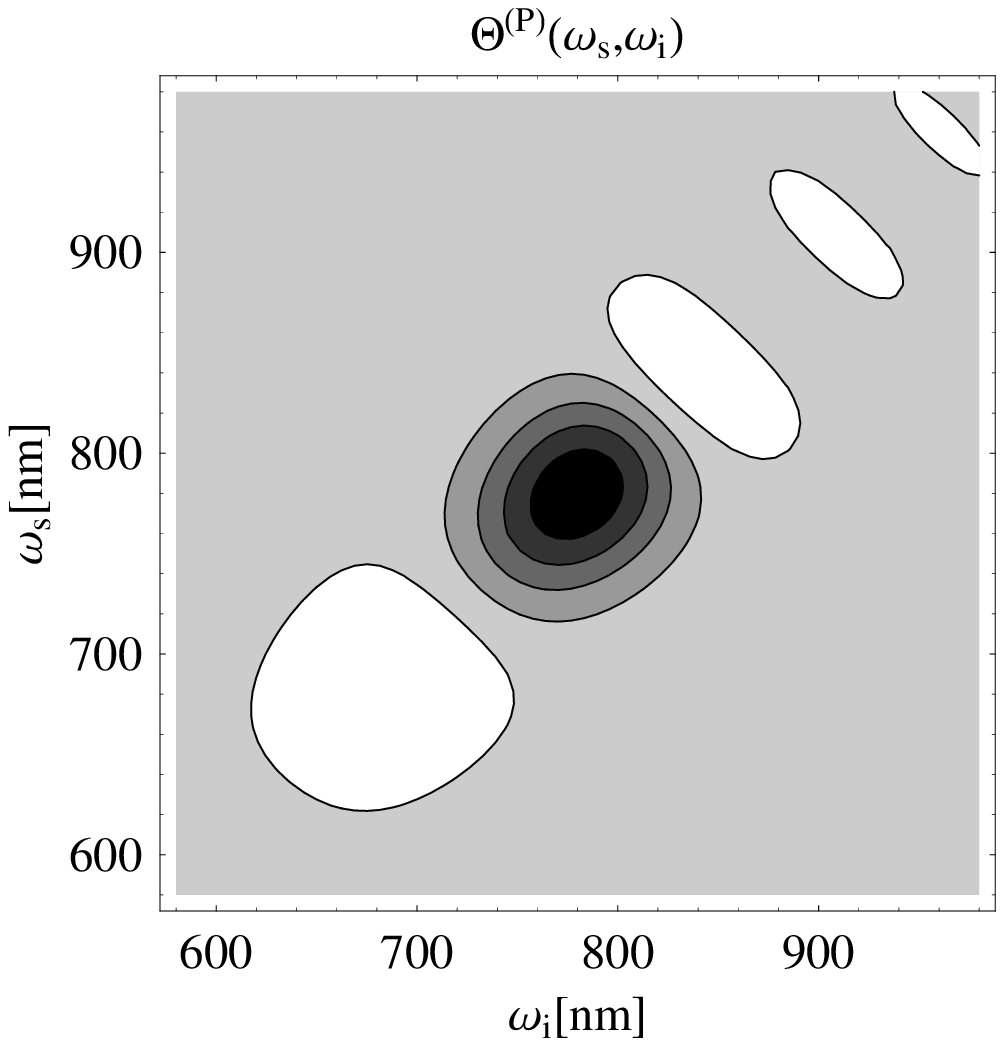}}&
  \subfigure[]{\includegraphics[width=0.49\columnwidth]{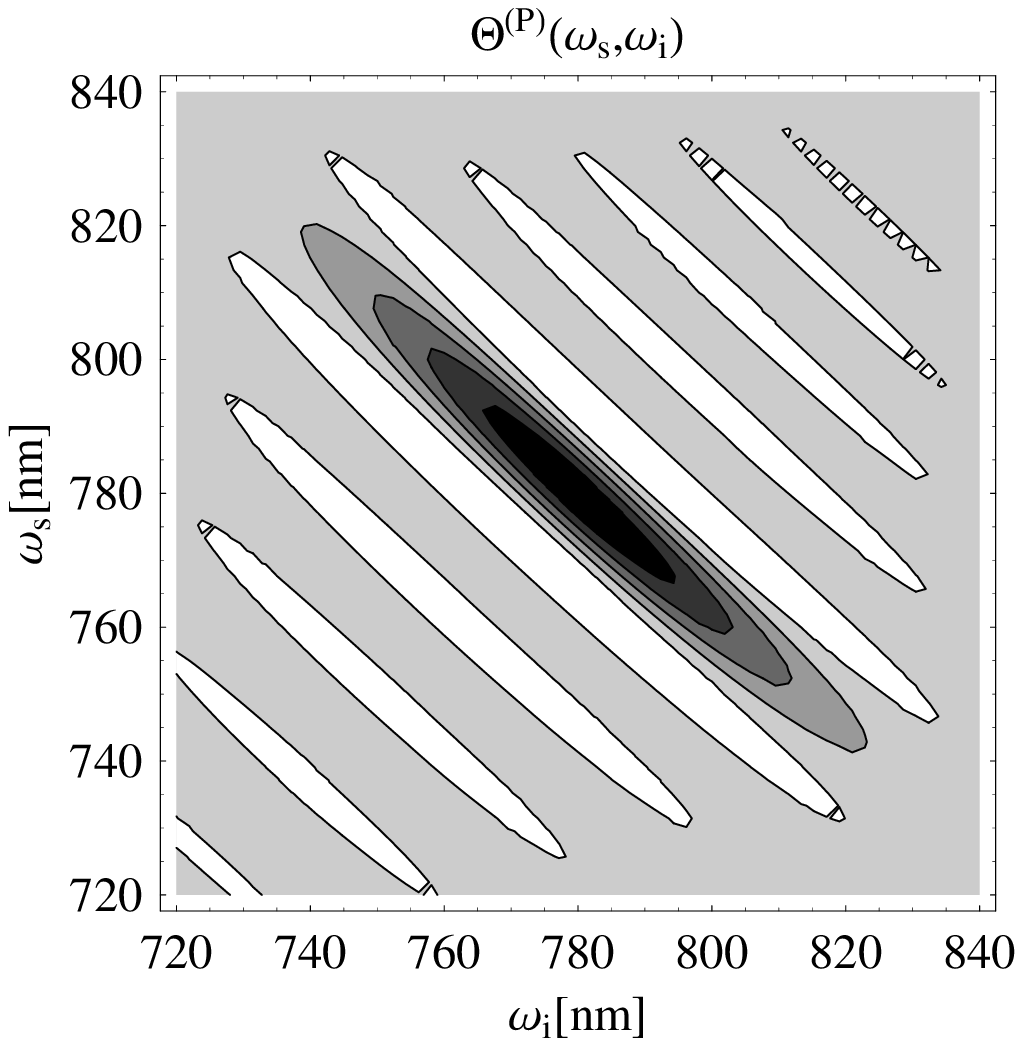}}\\
  \subfigure[]{\includegraphics[width=0.49\columnwidth]{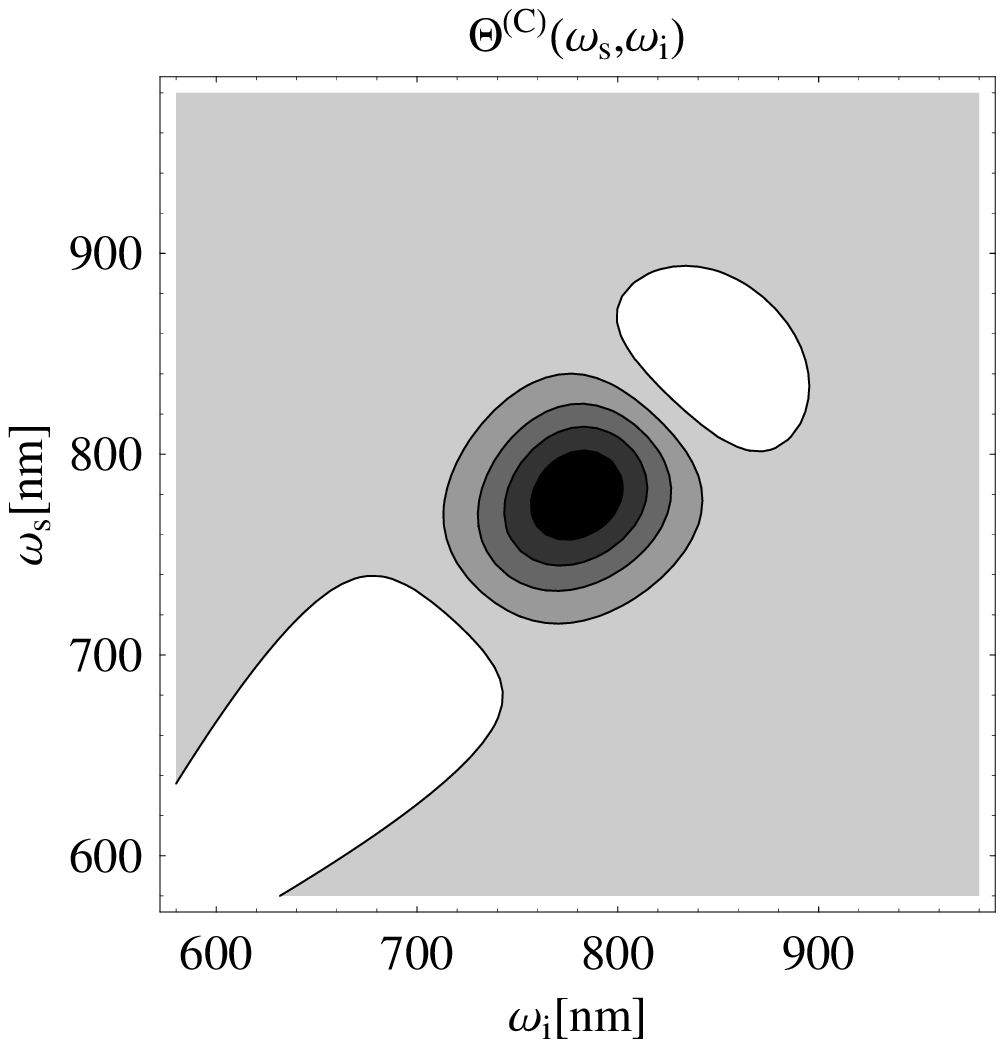}}&
  \subfigure[]{\includegraphics[width=0.49\columnwidth]{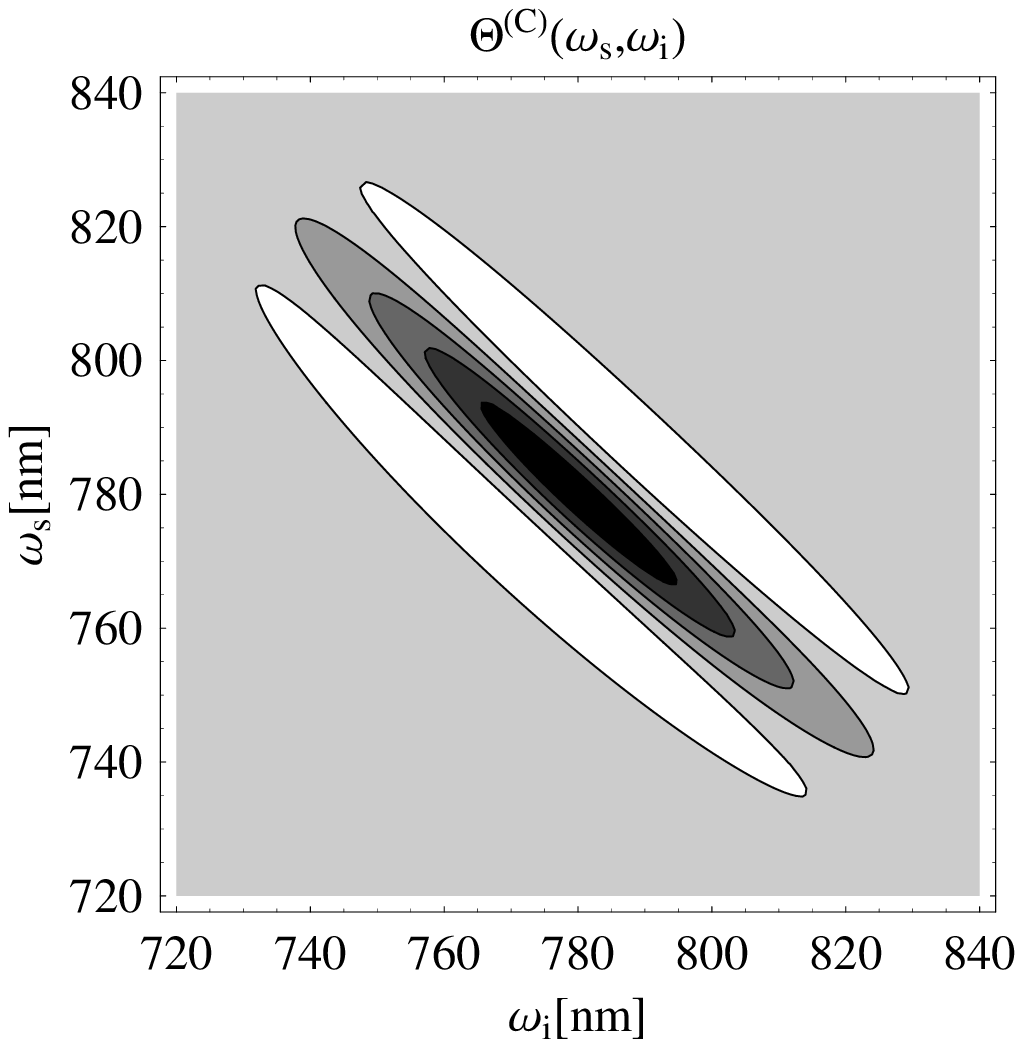}}\\
  \subfigure[]{\includegraphics[width=0.49\columnwidth]{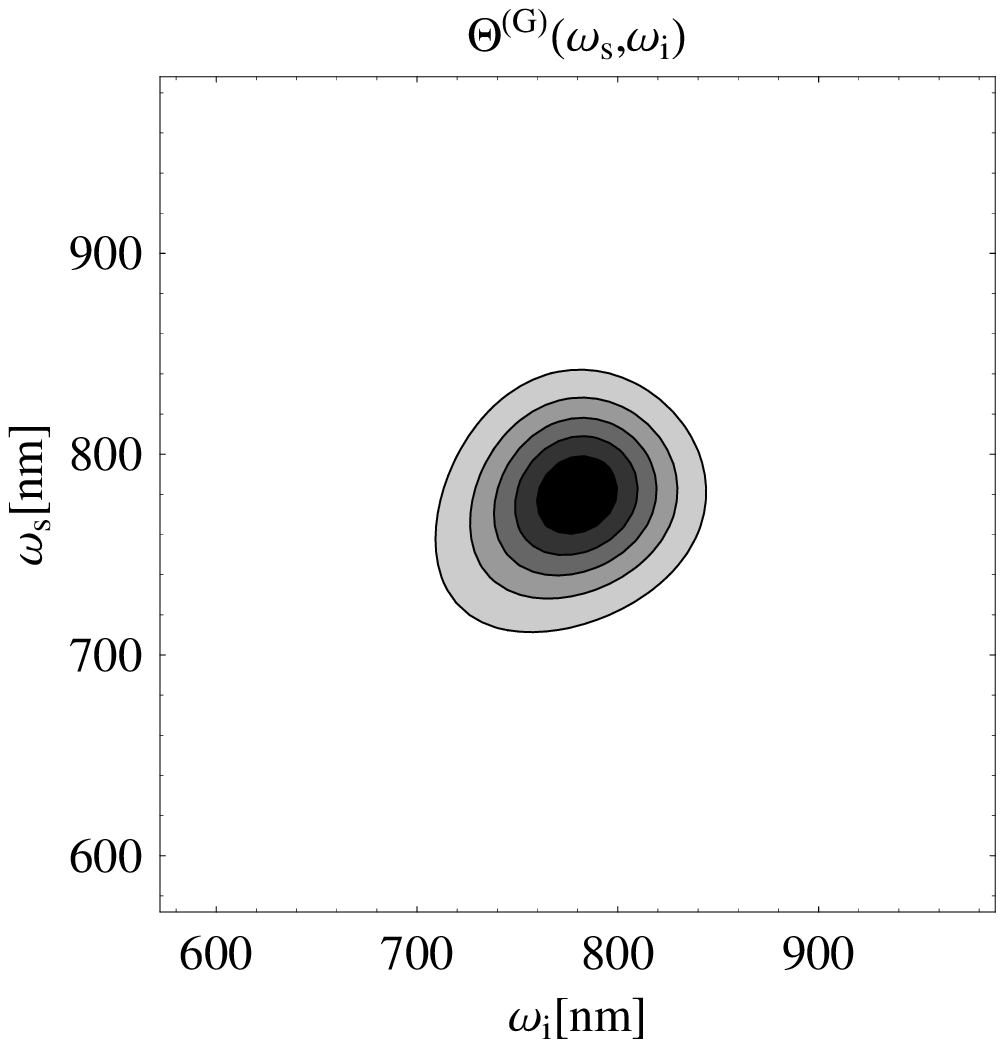}}&
  \subfigure[]{\includegraphics[width=0.49\columnwidth]{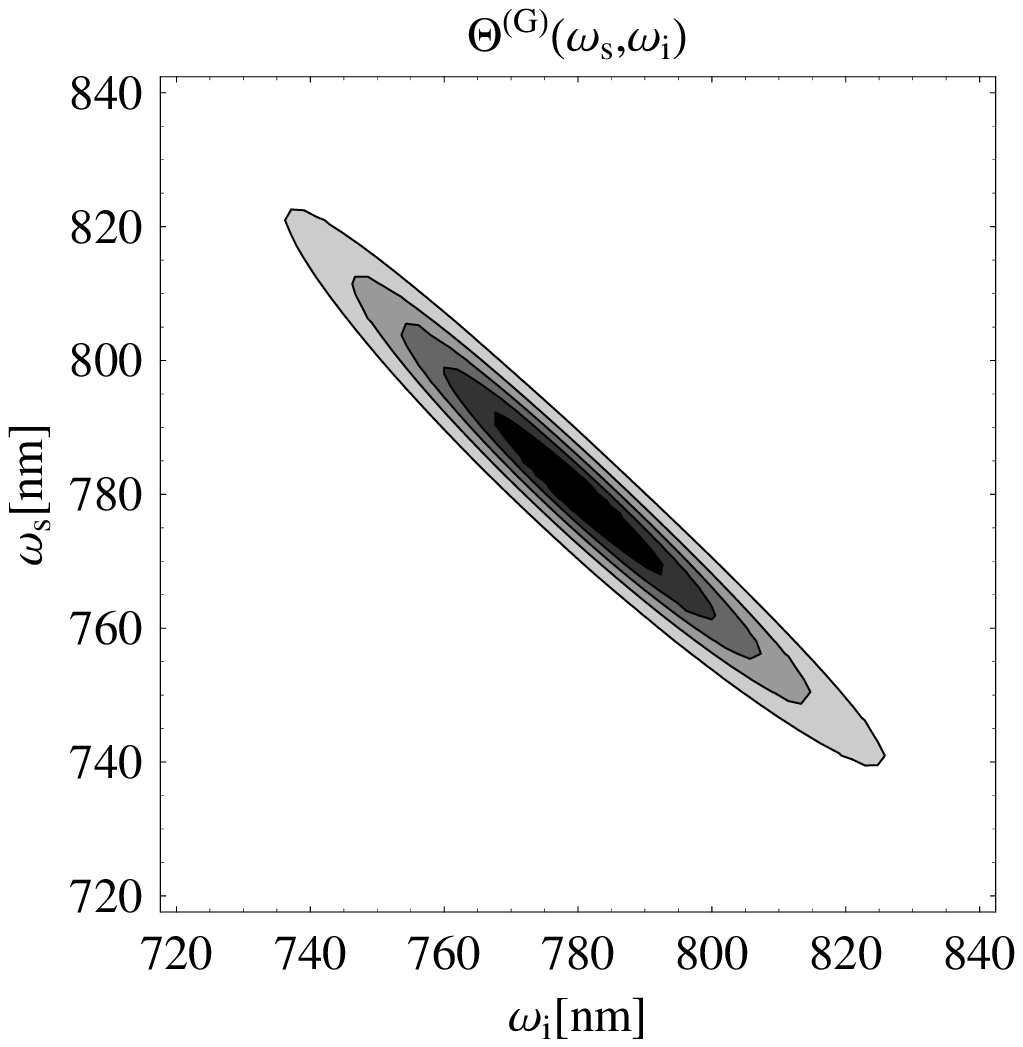}}\\
\end{tabular}
  \caption{The effective phase matching function $\Theta(\omega_s,\omega_i)$ calculated using (a, b) direct numerical integration; (c, d) paraxial approximation; (e, f) cosine-gaussian approximation and (g, h) gaussian approximation for the crystal length
  (a, c, e, g) $L=100~\mu$m and (b, d, f, h) $L=1$~mm. The pump and collecting beam waists were set to $w_s=w_i=2w_p=70~\mu$m. The angular frequencies $\omega_s$ and $\omega_i$ are labelled with the corresponding wavelengths.}
  \label{fig:epmf}
\end{figure}

Let us now compare computational methods introduced in the preceding sections for typical experimental settings. In \figref{fig:epmf} we depict the effective phase matching function $\Theta(\omega_s,\omega_i)$ for two exemplary lengths of the nonlinear medium calculated using direct numerical integration, the paraxial approximation,
the cosine-gaussian approximation and the gaussian approximation. Calculations were carried out for a beta-barium borate crystal with its optical axis lying in the plane of the collected modes and cut at $\theta_c=30^\circ$ with respect to $z$ axis. This corresponds to the symmetric cone half-opening angle equal to $\alpha=2.2^\circ$ for frequency-degenerate photons at $780$~nm. The beam waists were set to rather low values $w_s=w_i=2w_p=70~\mu$m to test the applicability limits of the paraxial approximation.

As seen in \figref{fig:epmf}, the main qualitative difference between the computational methods is the reproduction of the side lobes. The impact of the side lobes on observable quantities depends on the spectral width of the pump pulse. If the spectral bandwidth is narrower than the width of the central peak, then all the models can be expected to yield similar results. Because the characteristic width of $\Theta(\omega_s,\omega_i)$ along the axis $\omega_s = \omega_i$ decreases with a longer crystal length,
this regime corresponds to sufficiently narrow spectral bandwidths and short crystals. When leaving this regime, CGA can be expected to yield more accurate results in the intermediate regime compared to GA, as it reproduces correctly the lobes closest to the central peak.

\begin{figure}[h]
  \begin{tabular}{c}
     \subfigure{\includegraphics[width=0.98\columnwidth]{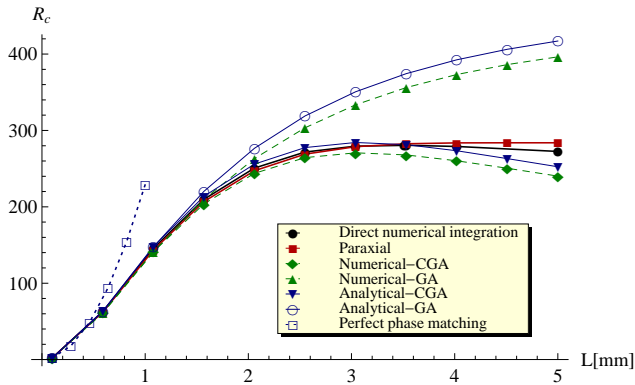}} \end{tabular}
  \caption{(Color online) The source brightness $R_c$ calculated using different numerical methods, specified in the inset, as a function of the crystal length $L$,  for the waists $w_s=w_i=2w_p=70~\mu$m and the pump pulse duration $\tp=100$~fs.}
  \label{fig:comaprisonRCRS}
\end{figure}

These predictions are confirmed by the calculation of the brightness $R_c$ as a function of the crystal length using different models, with the results shown in \figref{fig:comaprisonRCRS}. The full width at half maximum of the gaussian pump pulse was taken equal to $\tp=\tau_p \sqrt{\ln 2} = 100$~fs.
The brightness has been calculated through two-dimensional numerical integration of $|\Psi(\omega_s,\omega_i)|^2$ over the signal and the idler frequencies on a $32 \times 32$ square grid centered at $\omega_0$ for the relevant frequency range where wave function is nonzero. We have found that the further increase of grid density to $64\times 64$ did not change the results noticeably. In the paraxial approximation, the effective phase matching function $\thP$ was evaluated at each point of the grid using Gauss-Kronrod quadrature with three-digit precision. Results based on numerical integration of $|\Psi(\omega_s,\omega_i)|^2$ involving CGA and GA expressions for the effective phase matching function have been labeled respectively as \emph{numerical CGA} and \emph{numerical GA}.
In addition, we present results of applying a further simplification to CGA and GA, labeled as \emph{analytical CGA} and \emph{analytical GA}. The simplification consists in expanding
the functions $f(\omega_s,\omega_i)$ and $g(\omega_s,\omega_i)$ that appear in \eqref{eq:epmf:CGA} around the central frequency $\omega_0$ up to the second order and replacing $\Gamma(\omega_s,\omega_i)$ by its value at
$\omega_s=\omega_i=\omega_0$. After this expansion the squared absolute value of the biphoton wave function becomes a sum of three gaussian components and the integration over the frequencies $\omega_s$ and $\omega_i$ can be carried out analytically.

\figref{fig:comaprisonRCRS} shows that for short crystals all the models give similar results. Furthermore, in this regime the brightness $R_c$ exhibits quadratic dependence on the crystal length, which agrees with \eqref{eq:zero:rc} derived under the assumption of perfect phase matching. As expected, with an increasing crystal length the GA model departs earlier from the numerical results than the CGA model.

A more thorough way to compare the paraxial approximation with direct numerical integration is to evaluate two quantities: the scalar product between the normalized biphoton wave functions $\phP$ and $\phD$ obtained using both methods and the ratio of the corresponding pair production rates $\RcP/\RcD$. We carried out these calculations in an  unfavorable regime of a long crystal $L=2$~mm, ultrashort pump pulses $\tp=20$~fs,
and strong focusing $w_s=w_i=2w_p=40\mu$m. We found that both the quantities differed from one by less than $10^{-3}$. It should be noted that the computational effort required by the paraxial approximation was reduced in our calculations by $\sim 10^4$ compared to the direct numerical integration.

\begin{figure}[h]
  \begin{tabular}{cc}
      \subfigure[$L=100$~$\mu$m]{  \label{fig:Comparizon:Method:L01} \includegraphics[width=0.45\columnwidth]{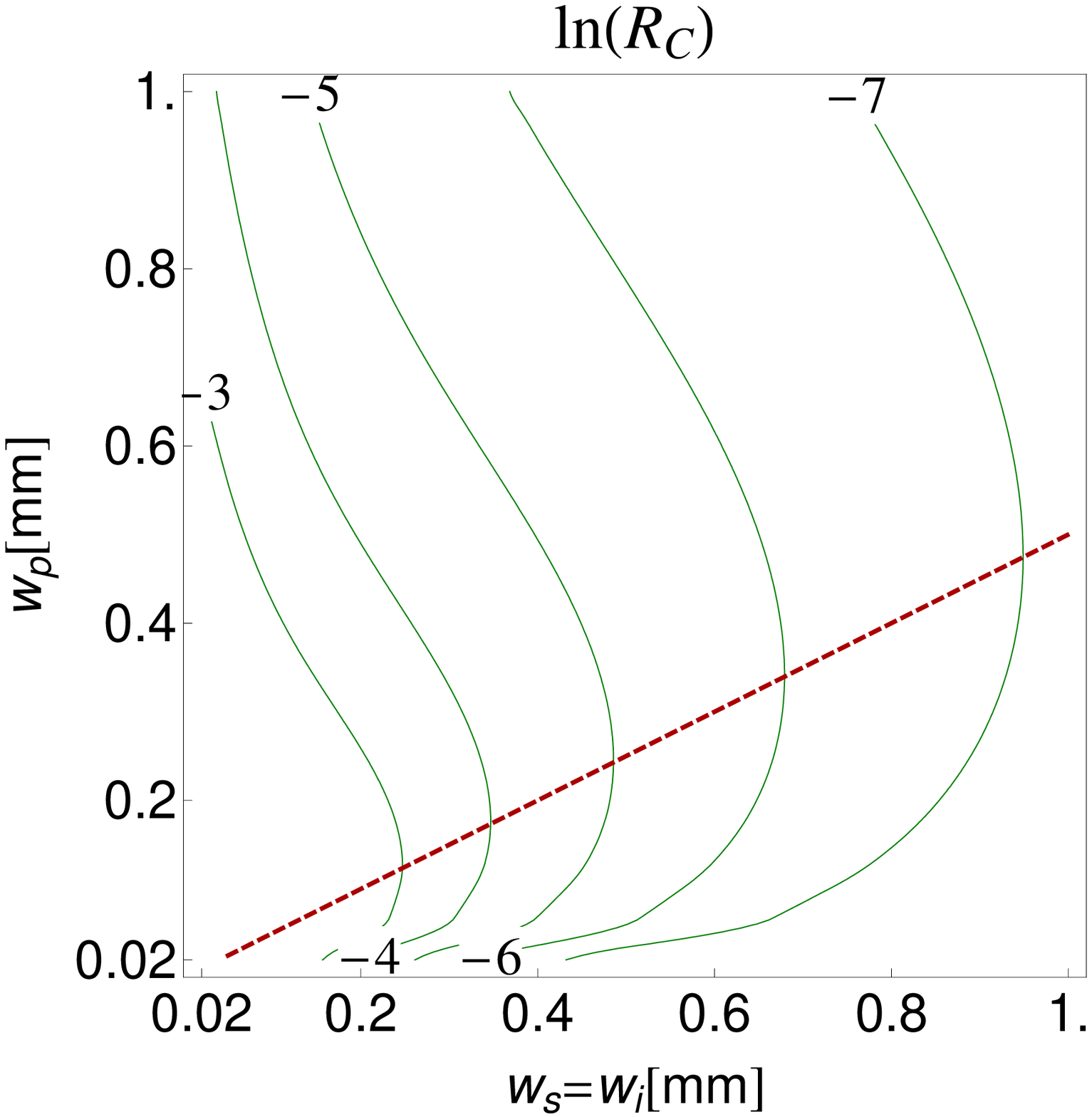}}&
      \subfigure[$L=1$~mm]{  \label{fig:Comparizon:Method:L2}
      \includegraphics[width=0.45\columnwidth]{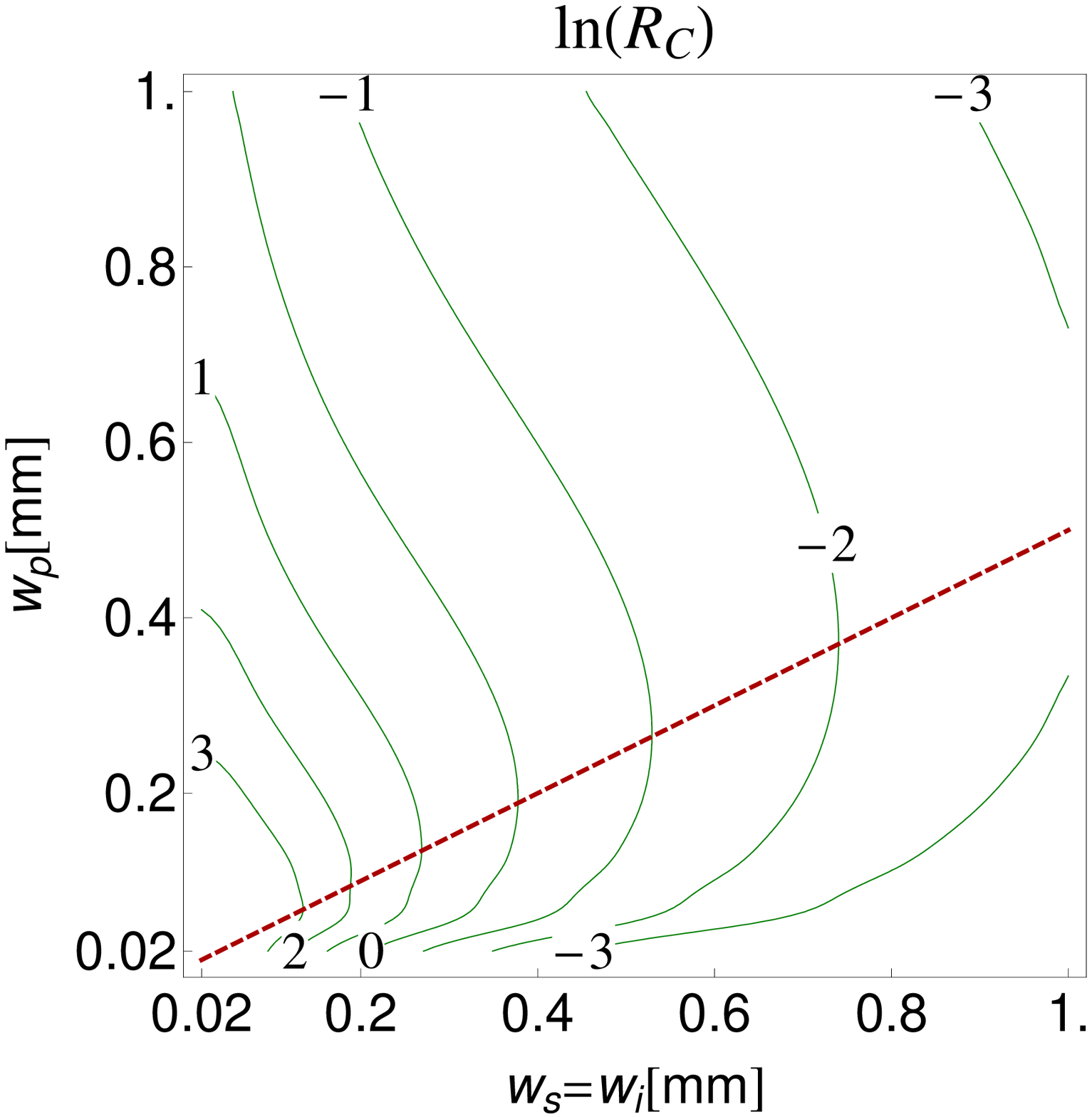}}
  \end{tabular}
  \caption{(Color online) The natural logarithm of the brightness $\ln R_c$ as a function of pump beam waist $w_p$ and fiber mode waists $w_s=w_i$ for the crystal length (a) $L=1$~mm and (b) $L=100~\mu$m.
  The dashed (red online) lines depict the condition $w_p= w_s/2$ specified in \eqref{eq:zero:optimalwaist}. }
  \label{fig:Comparizon:Method}
\end{figure}

Finally, let us analyze the coincidence count rate $R_c$ as a function of the pump beam waist $w_p$ and the fiber mode waists in a symmetric setup, when $w_s=w_i$. In \figref{fig:Comparizon:Method} we depict results obtained using the paraxial approximation for two exemplary lengths of the crystal. It is seen that for a fixed waist of the fiber modes the brightness has a well pronounced maximum in $w_p$. This maximum is located to a good approximation at $w_p=w_s/2$, which is in an agreement with the result derived within the elementary model of perfect phase matching in \eqref{eq:zero:optimalwaist}. This motivated the choice of $w_s=w_i=2w_p$ in the presented examples.

\section{Spectrally uncorrelated pairs}\label{section:SpectrallyUncorrelatedPairs}
A necessary condition for high-visibility multiphoton interference is the lack of distinguishing information about the origin of the photons, which means that each the photon should be prepared in an identical pure wavepacket. The most obvious way to achieve this regime is to insert interference filters whose bandwidth is smaller than the characteristic scale of spectral correlations within photon pairs. An intriguing alternative has been presented in Ref.~\cite{URen2005} which proposed to remove spectral correlations by exploiting geometric effects in SPDC. The purity of the produced photons needs to be analyzed in conjunction with other characteristics of the source, such as the pair production rate. In this section we will employ our computation tools to compare properties of spectrally decorrelated pairs generated by different methods.

%\subsection{Spectral decorrelation without interference filters}
Let us first analyze the geometric approach of Ref.~\cite{URen2005}. The underlying physics can be understood intuitively by looking at the biphoton wave function in the perfect phase matching approximation given by \eqref{eq:wf:zero}. The spectral pump amplitude introduces anticorrelations between frequencies of the down-converted photons, while the pump beam waist and emission angles define the degree of positive correlations. By balancing these two effects one can obtain a factorable biphoton wave function. More generally, without the approximation of perfect phase matching, one needs to analyze correlations introduced by the function $\Theta(\omega_s,\omega_i)$ defined in \eqref{eq:def:theta} combined with the spectral pump amplitude. As the nonlinear medium we considered a BBO crystal in the same configuration as discussed in Sec.~\ref{section:Comparison}. As the basic tool, we chose the paraxial method developed in \secref{section:NumericalApproach} due to its high precision and computational effectiveness. In order to evaluate the purity parameter $\pur$ measuring degree of spectral correlations, the approach presented by Law \emph{et al.}~\cite{law2000} was used. The method is based on the singular value decomposition of the matrix representation of the biphoton wave function $\Psi(\omega_s,\omega_i)$ on a sufficiently fine discrete grid. The normalized singular values are approximations of Schmidt coefficients $\varsigma_n$ and as such are used to evaluate purity parameter $\pur$. We found it sufficient to take the grid $32\times32$. Further increase of the grid density did not make any noticeable difference.
\begin{figure}[h]
  \centering
  \includegraphics[width=1\columnwidth]{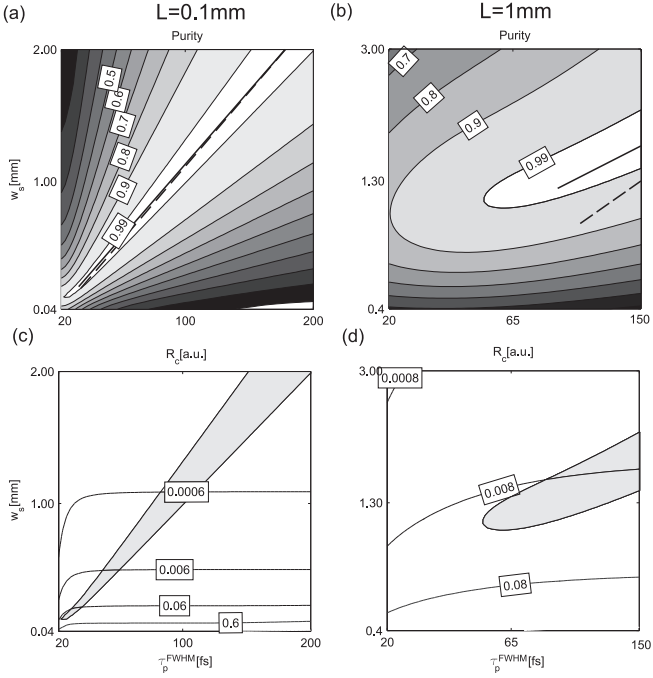}
  \caption{Contour plots of (a, b) the purity parameter $\pur$ and (c, d) the brightness $R_c$ as a function of the pump pulse duration $\tp$ and the collected mode waists $w_s=w_i$. The crystal thickness is (a, c) $L=1~\text{mm}$ and (b,d) $L=0.1$~mm. The solid and dashed lines in (a, b) correspond to factorability conditions given respectively in Eqs.~(\ref{eq:tpdec}) and (\ref{eq:zero:dec}). The grey areas in (c, d) mark the regions where the purity parameter is $\pur \ge 0.99$.}\label{fig:ksimap}
\end{figure}

In \figref{fig:ksimap}(a,b) we present the purity parameter for two typical lengths of the crystal as a function of the pulse duration $\tp$ and the collecting mode waist $w_s$. We assumed that the waists of the fiber modes and pump beam are $w_s = w_i = 2w_p$, which is motivated by the results presented in \figref{fig:Comparizon:Method}. The contour plots exhibit a clear relation between $\tp$ and $w_s$ that leads to minimized spectral correlations between photons. For a comparison, \figref{fig:ksimap}(a) and (b) depict also the purity condition derived in \eqref{eq:tpdec} using the GA model, as well as the predictions of the perfect phase matching approximation given in \eqref{eq:zero:dec}. It is seen that for the shorter crystal length $L=100$~$\mu$m the simple analytical formula of \eqref{eq:zero:dec} gives accurate results. This is because the spectral anticorrelations are predominantly defined by the pump bandwidth rather than the phase matching of the crystal. This is no longer valid for the length $L=1$~mm, where the effective bandwidth of the down-conversion process becomes strongly affected by the phase matching. These observations are consistent with results presented in \figref{fig:comaprisonRCRS}:  for $L=100$~$\mu$m the pair production rate is accurately given by the perfect phase matching approximation, while for $L=1$~mm effects of finite phase matching bandwidth are clearly seen.

The relation between the collecting mode waist $w_s=w_i$ and the pump pulse duration $\tp$ that leads to minimized spectral correlations gives us some flexibility to optimize the source with respect to other parameters. In \figref{fig:ksimap}(c,d) we present the source brightness $R_c$ as a function of $w_s$ and $\tau_p$. Note that in our calculations we constrain the pump beam waist by imposing $w_s=w_i=2w_p$. It is seen that $R_c$ can be increased by reducing the fiber mode waist $w_s$. As Figs.~\ref{fig:ksimap}(c) and \ref{fig:ksimap}(d) depict the pair production rate in the same units, we can compare the brightness for the two crystal lengths. Assuming that we have no restrictions on the pump pulse duration, a shorter crystal can produce more uncorrelated photon pairs. This is because for $L=1$~mm stronger spectral anticorrelations overwhelm the benefit of a longer nonlinear medium. However, in a realistic situation there is usually a technical minimum on the pump pulse duration. For concreteness, let us assume it to be
$\tp = 100$~fs. An inspection of \figref{fig:ksimap} shows that under the condition of nearly ideal decorrelation defined by the value of the purity parameter $\pur\approx 0.99$ higher brightness, approximately equal to $R_c \approx 0.046$, is obtained when the fiber mode waist is $w_s\simeq 1$~mm and the crystal length $L=1$~mm. We found that for even longer crystals decorrelation can be reached only using longer, less focused pump pulses, which lowers the source brightness.

%Do doktoratu
%\com{Czy dla $L=2$mm da sie w ogole dobic?} (We found that for longer crystals the purity parameter cannot be reached....)

\begin{figure}[ht]
\centering
\includegraphics[width=0.95\columnwidth]{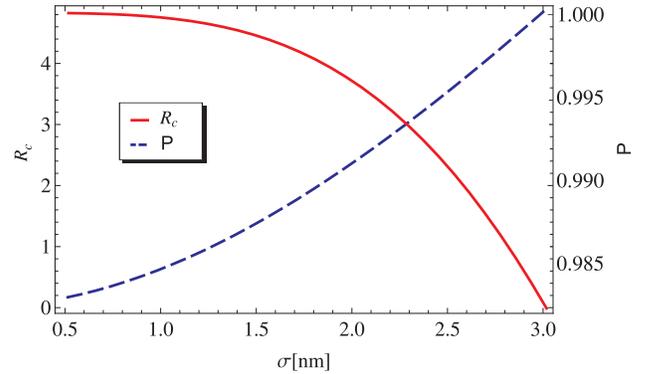}
\caption{(Color online) The brightness $R_c$ (dashed blue line, left vertical scale) and the purity parameter $\pur$ (solid red line, right vertical scale) as a function of the spectral filter bandwidth  $\sigma=\sigma_s=\sigma_i$ for a crystal length $L=1$mm, beam waists $w_s=w_i=2w_p =100\mu$m and the pump pulse duration $\tau_p=100$~fs.
}
\label{fig:StrongFocus}
\end{figure}

These limitations raise the question whether a more efficient strategy may rely on collecting tightly focused modes and removing spectral correlations with interference filters. Let us consider the same pump pulse duration $\tp = 100$~fs and crystal length $L=1$~mm as before, but tighten the fiber mode waists to $w_s=100~\mu$m.
The result is significantly increased brightness, but at the cost of introducing spectral correlations. The effects of inserting interference filters into such a setup are shown in \figref{fig:StrongFocus}, where we depict the brightness $R_c$ and
the purity parameter $\pur$ as a function of the spectral filter bandwidth. It is seen that for the bandwidth $\sigma \approx 2.6$~nm the purity parameter reaches the value $\pur \approx 0.99$, while the brightness is $R_c \approx 3.8$, which is significantly higher than before. Thus the benefit of increased brightness is retained despite spectral filtering.

\begin{figure}[h]
\centering
       \includegraphics[width=0.95\columnwidth]{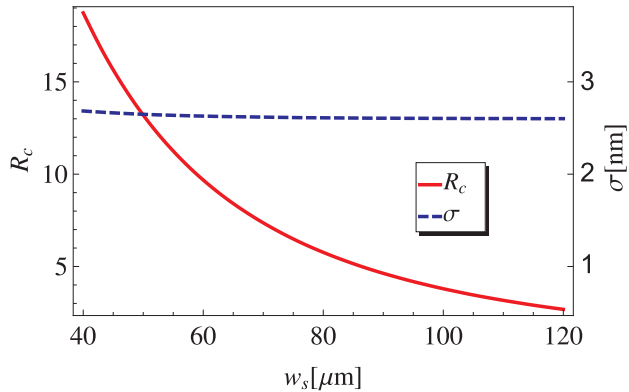}
\caption{(Color online)  The brightness $R_c$ (solid red line, left vertical scale) as a function of the collected mode waist $w_s$ obtained for the maximum filter bandwidth (dashed blue line, right vertical scale) which yields the purity parameter above $\pur \ge 0.99$. Other setup parameters are identical as in Fig.~\protect\ref{fig:StrongFocus}.}
\label{fig:ScanTpandWp}
\end{figure}

In order to gain more insight into the trade-off between the source brightness and spectral correlations, we calculated the maximum filter bandwidth that gives the purity $\pur \simeq 0.99$ for a range of pump beam waists $w_p$, while keeping other parameters of the setup identical as in previous examples. The results are shown in \figref{fig:ScanTpandWp}. It is seen that the filter bandwidth across the analyzed range does not deviate significantly from the value $\sigma \cong 2.7$~nm, while the brightness increases substantially with tighter focusing. This can be explained by the fact that the spectral filter bandwidth is defined by the requirement to remove frequency anticorrelations which depend primarily on the crystal length and the pump pulse duration rather than the beam waist.

%We also address a question what is the optimal spectral filter width for a given setup parameters. By optimality we mean such a width which on one hand assures the reduction of spectral correlations to $\pur > 0.99$ and on the other hand gives the highest pair production rate $R_c$.  Note that there is a tradeoff, because the first condition imposes tightening the spectral filter width and in order to fulfill the last one the bigger width is desired. For our example the optimal choice is the filter width $\sigma\cong2.2$nm corresponding to the intersection of green dashed dotted line with blue dashed curve see \figref{fig:StrongFocus}. In general the problem needs to be solved numerically. For the exemplary geometry introduced in section \secref{section:Comparison} we consider the optimal filter width in situation when pulse duration is fixed to $\tau_p=100$fs, see \figref{fig:ScanTpandWp}. We observe that the optimal filter width is in good approximation constant equal $\sigma \cong 2.7$nm. Thus in this situation placing the spectral filter one my optimizing by changing only pumping and collecting optics.

\section{Conclusions}
In this paper we introduced and utilized approximate methods that alleviate the numerical load necessary to model SPDC sources while retaining the accuracy of the results in physically relevant regimes. Our approach was based on an observation that optical fibers collecting photons effectively define a relatively narrow range of wave vectors that needs to be included in calculations. This justified applying the paraxial approximation, which made a substantial portion of the problem tractable analytically and significantly reduced the remaining numerical effort. The paraxial approximation can be also combined with a simplification of the two-photon wave function to an analytically manageable form that led to closed formulas. We exploited these strategies to analyze performance parameters that characterize the usefulness of SPDC sources for quantum information applications, such as the pair production rate and the spectral purity parameter that is critical in multiphoton interference experiments involving multiple sources.

The choice of a computation method depends on the range of the setup parameters. The most difficult regime to deal with is that of very broadband, tightly focused pump pulses and long crystals. It is then necessary to include with high precision the phase matching function over a wide range of frequencies and transverse wave vectors. The most universal method is then direct numerical integration, which however requires tremendous computational effort. In practical situations, the paraxial approximation, delivers highly accurate results with significantly reduced numerical load for typical setup parameters. The validity of the paraxial approximation can be checked with a relatively low effort by comparing it with direct numerical integration only at the edges of the region of interest that correspond to most unfavorable cases. Such a confirmation allows one to apply the paraxial approximation throughout the entire region of interest reducing the overall computational cost. For examples studied in Sec.~\ref{section:SpectrallyUncorrelatedPairs}, the paraxial approximation has been verified to yield results that did not differ by more than few percent from direct numerical integration. In more restricted scenarios, one may consider using the cosine-gaussian approximation, which extends the validity of the previously used gaussian approximation. Results obtained with these methods can be used as a starting point for designing source characteristics with more elaborate and precise tools. We also discussed a crude approximation of perfect phase matching, which gives simple, qualitative insights into the roles played by various source parameters.

The numerical methods presented in this work can be used to analyze various aspects of down-conversion sources that are relevant to experimental implementations of quantum information processing protocols. We discussed here spectral decorrelation, which is a necessary condition to achieve high-visibility multiphoton interference between independent sources, in connection with the pair production rate. For exemplary settings
chosen for the analysis, we found that spectral filtering combined with tight focusing of the pump beam can deliver higher brightness than balancing the spectral correlations using the geometry of the setup. The paraxial approximation can be also extended to analyze properties of an individual photon generated in the down-conversion process, with traced out degrees of freedom of the conjugate photon. This approach has been successfully applied to model the results of a measurement of the single-photon density matrix in the spectral domain reported in Ref.~\cite{wasilewski2007}. Theoretical details of this work will be presented elsewhere \cite{Kolenderski2009a}. Furthermore, the single photon count rates allow us to calculate the heralding efficiency, defined as the ratio of the pair production rate to the count rate on the trigger detector. This is another important parameter characterizing the usefulness of down-conversion sources \cite{Mosley2008}, that can be efficiently submitted to numerical optimization using paraxial approximation. We aim to make this a subject of a separate publication.
The numerical results presented in this paper have been obtained using a Mathematica code which can be downloaded from \footnote{{http://www.fizyka.umk.pl/$\sim$kolenderski/}}.

%
%\com{dopisac jeszcze}
%We used the developed tools to show for BBO crystal that it is possible to adjust the source parameters in order to produce fiber coupled photon pairs with reduced spectral correlation without the need of use the spectral filters. This method however is inefficient, because of low fiber coupled pair production rate $R_c$, compared to presented alternative method in which the strong focusing of pump beam along with the spectral filtering is employed. Nevertheless we showed the method how to choose the optimal spectral filter for the configuration with strong pump focusing in order to get both high fiber coupled pair production rate and reduced spectral correlations.

\section{Acknowledgements}
PK acknowledges the insightful conversation with Jan Iwaniszewski. This work has been supported by Polish MNISW (N N202 1489 33) and the European Commission under the Integrated Project Qubit Applications (QAP) funded by the IST directorate as Contract Number 015848.

%\bibliography{PKbase}

\begin{thebibliography}{32}
\expandafter\ifx\csname natexlab\endcsname\relax\def\natexlab#1{#1}\fi
\expandafter\ifx\csname bibnamefont\endcsname\relax
  \def\bibnamefont#1{#1}\fi
\expandafter\ifx\csname bibfnamefont\endcsname\relax
  \def\bibfnamefont#1{#1}\fi
\expandafter\ifx\csname citenamefont\endcsname\relax
  \def\citenamefont#1{#1}\fi
\expandafter\ifx\csname url\endcsname\relax
  \def\url#1{\texttt{#1}}\fi
\expandafter\ifx\csname urlprefix\endcsname\relax\def\urlprefix{URL }\fi
\providecommand{\bibinfo}[2]{#2}
\providecommand{\eprint}[2][]{\url{#2}}

\bibitem[{\citenamefont{Clauser et~al.}(1969)\citenamefont{Clauser, Horne,
  Shimony, and Holt}}]{CHSH1969}
\bibinfo{author}{\bibfnamefont{J.~F.} \bibnamefont{Clauser}},
  \bibinfo{author}{\bibfnamefont{M.~A.} \bibnamefont{Horne}},
  \bibinfo{author}{\bibfnamefont{A.}~\bibnamefont{Shimony}}, \bibnamefont{and}
  \bibinfo{author}{\bibfnamefont{R.~A.} \bibnamefont{Holt}},
  \bibinfo{journal}{Phys. Rev. Lett.} \textbf{\bibinfo{volume}{23}},
  \bibinfo{pages}{880} (\bibinfo{year}{1969}).

\bibitem[{\citenamefont{Kwiat et~al.}(1995)\citenamefont{Kwiat, Mattle,
  Weinfurter, Zeilinger, Sergienko, and Shih}}]{Kwiat1995}
\bibinfo{author}{\bibfnamefont{P.~G.} \bibnamefont{Kwiat}},
  \bibinfo{author}{\bibfnamefont{K.}~\bibnamefont{Mattle}},
  \bibinfo{author}{\bibfnamefont{H.}~\bibnamefont{Weinfurter}},
  \bibinfo{author}{\bibfnamefont{A.}~\bibnamefont{Zeilinger}},
  \bibinfo{author}{\bibfnamefont{A.~V.} \bibnamefont{Sergienko}},
  \bibnamefont{and} \bibinfo{author}{\bibfnamefont{Y.}~\bibnamefont{Shih}},
  \bibinfo{journal}{Phys. Rev. Lett.} \textbf{\bibinfo{volume}{75}},
  \bibinfo{pages}{4337} (\bibinfo{year}{1995}).

\bibitem[{\citenamefont{Boschi et~al.}(1998)\citenamefont{Boschi, Branca,
  De~Martini, Hardy, and Popescu}}]{Boschi1998}
\bibinfo{author}{\bibfnamefont{D.}~\bibnamefont{Boschi}},
  \bibinfo{author}{\bibfnamefont{S.}~\bibnamefont{Branca}},
  \bibinfo{author}{\bibfnamefont{F.}~\bibnamefont{De~Martini}},
  \bibinfo{author}{\bibfnamefont{L.}~\bibnamefont{Hardy}}, \bibnamefont{and}
  \bibinfo{author}{\bibfnamefont{S.}~\bibnamefont{Popescu}},
  \bibinfo{journal}{Phys. Rev. Lett.} \textbf{\bibinfo{volume}{80}},
  \bibinfo{pages}{1121} (\bibinfo{year}{1998}).

\bibitem[{\citenamefont{Marcikic et~al.}(2003)\citenamefont{Marcikic,
  de~Riedmatten, Tittel, Zbinden, and Gisin}}]{Marcikic2003}
\bibinfo{author}{\bibfnamefont{I.}~\bibnamefont{Marcikic}},
  \bibinfo{author}{\bibfnamefont{H.}~\bibnamefont{de~Riedmatten}},
  \bibinfo{author}{\bibfnamefont{W.}~\bibnamefont{Tittel}},
  \bibinfo{author}{\bibfnamefont{H.}~\bibnamefont{Zbinden}}, \bibnamefont{and}
  \bibinfo{author}{\bibfnamefont{N.}~\bibnamefont{Gisin}},
  \bibinfo{journal}{Nature} \textbf{\bibinfo{volume}{421}},
  \bibinfo{pages}{509} (\bibinfo{year}{2003}).

\bibitem[{\citenamefont{Ursin et~al.}(2004)\citenamefont{Ursin, Jennewein,
  Aspelmeyer, Kaltenbaek, Lindenthal, Walther, and Zeilinger}}]{Ursin2004}
\bibinfo{author}{\bibfnamefont{R.}~\bibnamefont{Ursin}},
  \bibinfo{author}{\bibfnamefont{T.}~\bibnamefont{Jennewein}},
  \bibinfo{author}{\bibfnamefont{M.}~\bibnamefont{Aspelmeyer}},
  \bibinfo{author}{\bibfnamefont{R.}~\bibnamefont{Kaltenbaek}},
  \bibinfo{author}{\bibfnamefont{M.}~\bibnamefont{Lindenthal}},
  \bibinfo{author}{\bibfnamefont{P.}~\bibnamefont{Walther}}, \bibnamefont{and}
  \bibinfo{author}{\bibfnamefont{A.}~\bibnamefont{Zeilinger}},
  \bibinfo{journal}{Nature} \textbf{\bibinfo{volume}{430}},
  \bibinfo{pages}{849} (\bibinfo{year}{2004}).

\bibitem[{\citenamefont{Gisin et~al.}(2002)\citenamefont{Gisin, Ribordy,
  Tittel, and Zbinden}}]{Gisin2002}
\bibinfo{author}{\bibfnamefont{N.}~\bibnamefont{Gisin}},
  \bibinfo{author}{\bibfnamefont{G.}~\bibnamefont{Ribordy}},
  \bibinfo{author}{\bibfnamefont{W.}~\bibnamefont{Tittel}}, \bibnamefont{and}
  \bibinfo{author}{\bibfnamefont{H.}~\bibnamefont{Zbinden}},
  \bibinfo{journal}{Rev. Mod. Phys.} \textbf{\bibinfo{volume}{74}},
  \bibinfo{pages}{145} (\bibinfo{year}{2002}).

\bibitem[{\citenamefont{Kok et~al.}(2007)\citenamefont{Kok, Munro, Nemoto,
  Ralph, Dowling, and Milburn}}]{Kok2007}
\bibinfo{author}{\bibfnamefont{P.}~\bibnamefont{Kok}},
  \bibinfo{author}{\bibfnamefont{W.~J.} \bibnamefont{Munro}},
  \bibinfo{author}{\bibfnamefont{K.}~\bibnamefont{Nemoto}},
  \bibinfo{author}{\bibfnamefont{T.~C.} \bibnamefont{Ralph}},
  \bibinfo{author}{\bibfnamefont{J.~P.} \bibnamefont{Dowling}},
  \bibnamefont{and} \bibinfo{author}{\bibfnamefont{G.~J.}
  \bibnamefont{Milburn}}, \bibinfo{journal}{Rev. Mod. Phys.}
  \textbf{\bibinfo{volume}{79}}, \bibinfo{eid}{135} (\bibinfo{year}{2007}).

\bibitem[{\citenamefont{Kaltenbaek et~al.}(2006)\citenamefont{Kaltenbaek,
  Blauensteiner, \.{Z}ukowski, Aspelmeyer, and Zeilinger}}]{Kaltenbaek2006}
\bibinfo{author}{\bibfnamefont{R.}~\bibnamefont{Kaltenbaek}},
  \bibinfo{author}{\bibfnamefont{B.}~\bibnamefont{Blauensteiner}},
  \bibinfo{author}{\bibfnamefont{M.}~\bibnamefont{\.{Z}ukowski}},
  \bibinfo{author}{\bibfnamefont{M.}~\bibnamefont{Aspelmeyer}},
  \bibnamefont{and}
  \bibinfo{author}{\bibfnamefont{A.}~\bibnamefont{Zeilinger}},
  \bibinfo{journal}{Phys.~Rev.~Lett.} \textbf{\bibinfo{volume}{96}},
  \bibinfo{eid}{240502} (\bibinfo{year}{2006}).

\bibitem[{\citenamefont{Riedmatten et~al.}(2003)\citenamefont{Riedmatten,
  Marcikic, Tittel, Zbinden, and Gisin}}]{Riedmatten2003}
\bibinfo{author}{\bibfnamefont{H.~d.} \bibnamefont{Riedmatten}},
  \bibinfo{author}{\bibfnamefont{I.}~\bibnamefont{Marcikic}},
  \bibinfo{author}{\bibfnamefont{W.}~\bibnamefont{Tittel}},
  \bibinfo{author}{\bibfnamefont{H.}~\bibnamefont{Zbinden}}, \bibnamefont{and}
  \bibinfo{author}{\bibfnamefont{N.}~\bibnamefont{Gisin}},
  \bibinfo{journal}{Phys. Rev. A} \textbf{\bibinfo{volume}{67}},
  \bibinfo{pages}{022301} (\bibinfo{year}{2003}).

\bibitem[{\citenamefont{Dragan}(2004)}]{Dragan2004}
\bibinfo{author}{\bibfnamefont{A.}~\bibnamefont{Dragan}},
  \bibinfo{journal}{Phys. Rev. A} \textbf{\bibinfo{volume}{70}},
  \bibinfo{eid}{053814} (\bibinfo{year}{2004}).

\bibitem[{\citenamefont{U'Ren et~al.}(2005)\citenamefont{U'Ren, Silberhorn,
  Banaszek, Walmsley, Erdmann, Grice, and Raymer}}]{URen2005}
\bibinfo{author}{\bibfnamefont{A.~B.} \bibnamefont{U'Ren}},
  \bibinfo{author}{\bibfnamefont{C.}~\bibnamefont{Silberhorn}},
  \bibinfo{author}{\bibfnamefont{K.}~\bibnamefont{Banaszek}},
  \bibinfo{author}{\bibfnamefont{I.~A.} \bibnamefont{Walmsley}},
  \bibinfo{author}{\bibfnamefont{R.}~\bibnamefont{Erdmann}},
  \bibinfo{author}{\bibfnamefont{W.~P.} \bibnamefont{Grice}}, \bibnamefont{and}
  \bibinfo{author}{\bibfnamefont{M.~G.} \bibnamefont{Raymer}},
  \bibinfo{journal}{Las. Phys.} \textbf{\bibinfo{volume}{15}},
  \bibinfo{pages}{1} (\bibinfo{year}{2005}), \eprint{quant-ph/0611019}.

\bibitem[{\citenamefont{U'Ren et~al.}(2007)\citenamefont{U'Ren,
  Jeronimo-Moreno, and Garcia-Gracia}}]{URen2007}
\bibinfo{author}{\bibfnamefont{A.~B.} \bibnamefont{U'Ren}},
  \bibinfo{author}{\bibfnamefont{Y.}~\bibnamefont{Jeronimo-Moreno}},
  \bibnamefont{and}
  \bibinfo{author}{\bibfnamefont{H.}~\bibnamefont{Garcia-Gracia}},
  \bibinfo{journal}{Phys. Rev. A} \textbf{\bibinfo{volume}{75}},
  \bibinfo{eid}{023810} (\bibinfo{year}{2007}).

\bibitem[{\citenamefont{Mosley et~al.}(2008)\citenamefont{Mosley, Lundeen,
  Smith, Wasylczyk, U'Ren, Silberhorn, and Walmsley}}]{Mosley2008}
\bibinfo{author}{\bibfnamefont{P.~J.} \bibnamefont{Mosley}},
  \bibinfo{author}{\bibfnamefont{J.~S.} \bibnamefont{Lundeen}},
  \bibinfo{author}{\bibfnamefont{B.~J.} \bibnamefont{Smith}},
  \bibinfo{author}{\bibfnamefont{P.}~\bibnamefont{Wasylczyk}},
  \bibinfo{author}{\bibfnamefont{A.~B.} \bibnamefont{U'Ren}},
  \bibinfo{author}{\bibfnamefont{C.}~\bibnamefont{Silberhorn}},
  \bibnamefont{and} \bibinfo{author}{\bibfnamefont{I.~A.}
  \bibnamefont{Walmsley}}, \bibinfo{journal}{Phys. Rev. Lett.}
  \textbf{\bibinfo{volume}{100}}, \bibinfo{eid}{133601} (\bibinfo{year}{2008}).

\bibitem[{\citenamefont{Kurtsiefer et~al.}(2001)\citenamefont{Kurtsiefer,
  Oberparleiter, and Weinfurter}}]{Kurtsiefer2001}
\bibinfo{author}{\bibfnamefont{C.}~\bibnamefont{Kurtsiefer}},
  \bibinfo{author}{\bibfnamefont{M.}~\bibnamefont{Oberparleiter}},
  \bibnamefont{and}
  \bibinfo{author}{\bibfnamefont{H.}~\bibnamefont{Weinfurter}},
  \bibinfo{journal}{Phys. Rev. A} \textbf{\bibinfo{volume}{64}},
  \bibinfo{pages}{023802} (\bibinfo{year}{2001}).

\bibitem[{\citenamefont{Bovino et~al.}(2003)\citenamefont{Bovino, Varisco,
  Maria~Colla, Castagnoli, Di~Giuseppe, and Sergienko}}]{Bovino2003}
\bibinfo{author}{\bibfnamefont{F.~A.} \bibnamefont{Bovino}},
  \bibinfo{author}{\bibfnamefont{P.}~\bibnamefont{Varisco}},
  \bibinfo{author}{\bibfnamefont{A.}~\bibnamefont{Maria~Colla}},
  \bibinfo{author}{\bibfnamefont{G.}~\bibnamefont{Castagnoli}},
  \bibinfo{author}{\bibfnamefont{G.}~\bibnamefont{Di~Giuseppe}},
  \bibnamefont{and} \bibinfo{author}{\bibfnamefont{A.~V.}
  \bibnamefont{Sergienko}}, \bibinfo{journal}{Opt. Commmun.}
  \textbf{\bibinfo{volume}{227}}, \bibinfo{pages}{343} (\bibinfo{year}{2003}).

\bibitem[{\citenamefont{Castelletto et~al.}(2004)\citenamefont{Castelletto,
  Degiovanni, Migdall, and Ware}}]{Castelletto2004}
\bibinfo{author}{\bibfnamefont{S.}~\bibnamefont{Castelletto}},
  \bibinfo{author}{\bibfnamefont{I.~P.} \bibnamefont{Degiovanni}},
  \bibinfo{author}{\bibfnamefont{A.}~\bibnamefont{Migdall}}, \bibnamefont{and}
  \bibinfo{author}{\bibfnamefont{M.}~\bibnamefont{Ware}}, \bibinfo{journal}{New
  J. of Phys} \textbf{\bibinfo{volume}{6}}, \bibinfo{pages}{87}
  (\bibinfo{year}{2004}).

\bibitem[{\citenamefont{Castelletto et~al.}(2005)\citenamefont{Castelletto,
  Castelletto, Degiovanni, Furno, Schettini, Migdall, and
  Ware}}]{Castelletto2005}
\bibinfo{author}{\bibfnamefont{S.}~\bibnamefont{Castelletto}},
  \bibinfo{author}{\bibfnamefont{S.}~\bibnamefont{Castelletto}},
  \bibinfo{author}{\bibfnamefont{I.}~\bibnamefont{Degiovanni}},
  \bibinfo{author}{\bibfnamefont{G.}~\bibnamefont{Furno}},
  \bibinfo{author}{\bibfnamefont{V.}~\bibnamefont{Schettini}},
  \bibinfo{author}{\bibfnamefont{A.}~\bibnamefont{Migdall}}, \bibnamefont{and}
  \bibinfo{author}{\bibfnamefont{M.}~\bibnamefont{Ware}},
  \bibinfo{journal}{IEEE Trans. Instr. Meas.} \textbf{\bibinfo{volume}{54}},
  \bibinfo{pages}{890 } (\bibinfo{year}{2005}).

\bibitem[{\citenamefont{Andrews et~al.}(2004)\citenamefont{Andrews, Pike, and
  Sarkar}}]{Andrews2004}
\bibinfo{author}{\bibfnamefont{R.}~\bibnamefont{Andrews}},
  \bibinfo{author}{\bibfnamefont{E.}~\bibnamefont{Pike}}, \bibnamefont{and}
  \bibinfo{author}{\bibfnamefont{S.}~\bibnamefont{Sarkar}},
  \bibinfo{journal}{Opt. Express} \textbf{\bibinfo{volume}{12}},
  \bibinfo{pages}{3264} (\bibinfo{year}{2004}).

\bibitem[{\citenamefont{Lee et~al.}(2005)\citenamefont{Lee, van Exter, and
  Woerdman}}]{Lee2005}
\bibinfo{author}{\bibfnamefont{P.~S.~K.} \bibnamefont{Lee}},
  \bibinfo{author}{\bibfnamefont{M.~P.} \bibnamefont{van Exter}},
  \bibnamefont{and} \bibinfo{author}{\bibfnamefont{J.~P.}
  \bibnamefont{Woerdman}}, \bibinfo{journal}{Phys. Rev. A}
  \textbf{\bibinfo{volume}{72}}, \bibinfo{eid}{033803} (\bibinfo{year}{2005}).

\bibitem[{\citenamefont{Ljunggren and Tengner}(2005)}]{ljunggren2005}
\bibinfo{author}{\bibfnamefont{D.}~\bibnamefont{Ljunggren}} \bibnamefont{and}
  \bibinfo{author}{\bibfnamefont{M.}~\bibnamefont{Tengner}},
  \bibinfo{journal}{Phys. Rev. A} \textbf{\bibinfo{volume}{72}},
  \bibinfo{eid}{062301} (\bibinfo{year}{2005}).

\bibitem[{\citenamefont{Ljunggren et~al.}(2006)\citenamefont{Ljunggren,
  Tengner, Marsden, and Pelton}}]{ljunggren2006}
\bibinfo{author}{\bibfnamefont{D.}~\bibnamefont{Ljunggren}},
  \bibinfo{author}{\bibfnamefont{M.}~\bibnamefont{Tengner}},
  \bibinfo{author}{\bibfnamefont{P.}~\bibnamefont{Marsden}}, \bibnamefont{and}
  \bibinfo{author}{\bibfnamefont{M.}~\bibnamefont{Pelton}},
  \bibinfo{journal}{Phys. Rev. A} \textbf{\bibinfo{volume}{73}},
  \bibinfo{eid}{032326} (\bibinfo{year}{2006}).

\bibitem[{\citenamefont{Ling et~al.}(2008)\citenamefont{Ling, Lamas-Linares,
  and Kurtsiefer}}]{Ling2008}
\bibinfo{author}{\bibfnamefont{A.}~\bibnamefont{Ling}},
  \bibinfo{author}{\bibfnamefont{A.}~\bibnamefont{Lamas-Linares}},
  \bibnamefont{and}
  \bibinfo{author}{\bibfnamefont{C.}~\bibnamefont{Kurtsiefer}},
  \bibinfo{journal}{Phys. Rev. A} \textbf{\bibinfo{volume}{77}},
  \bibinfo{pages}{043834} (\bibinfo{year}{2008}).

\bibitem[{\citenamefont{Wasilewski et~al.}(2006)\citenamefont{Wasilewski,
  Wasylczyk, Kolenderski, Banaszek, and Radzewicz}}]{Wasilewski2006}
\bibinfo{author}{\bibfnamefont{W.}~\bibnamefont{Wasilewski}},
  \bibinfo{author}{\bibfnamefont{P.}~\bibnamefont{Wasylczyk}},
  \bibinfo{author}{\bibfnamefont{P.}~\bibnamefont{Kolenderski}},
  \bibinfo{author}{\bibfnamefont{K.}~\bibnamefont{Banaszek}}, \bibnamefont{and}
  \bibinfo{author}{\bibfnamefont{C.}~\bibnamefont{Radzewicz}},
  \bibinfo{journal}{Opt. Lett.} \textbf{\bibinfo{volume}{31}}
  (\bibinfo{year}{2006}), \eprint{quant-ph/0512039}.

\bibitem[{\citenamefont{Wasilewski et~al.}(2007)\citenamefont{Wasilewski,
  Kolenderski, and Frankowski}}]{wasilewski2007}
\bibinfo{author}{\bibfnamefont{W.}~\bibnamefont{Wasilewski}},
  \bibinfo{author}{\bibfnamefont{P.}~\bibnamefont{Kolenderski}},
  \bibnamefont{and}
  \bibinfo{author}{\bibfnamefont{R.}~\bibnamefont{Frankowski}},
  \bibinfo{journal}{Phys. Rev. Lett.} \textbf{\bibinfo{volume}{99}},
  \bibinfo{eid}{123601} (\bibinfo{year}{2007}).

\bibitem[{\citenamefont{U'Ren et~al.}(2003)\citenamefont{U'Ren, Banaszek, and
  Walmsley}}]{URen2003}
\bibinfo{author}{\bibfnamefont{A.}~\bibnamefont{U'Ren}},
  \bibinfo{author}{\bibfnamefont{K.}~\bibnamefont{Banaszek}}, \bibnamefont{and}
  \bibinfo{author}{\bibfnamefont{I.}~\bibnamefont{Walmsley}},
  \bibinfo{journal}{Quantum Infor. Compt.} \textbf{\bibinfo{volume}{3}},
  \bibinfo{pages}{480} (\bibinfo{year}{2003}), \eprint{quant-ph/0305192}.

\bibitem[{\citenamefont{Louisell et~al.}(1961)\citenamefont{Louisell, Yariv,
  and Siegman}}]{Louisell1961}
\bibinfo{author}{\bibfnamefont{W.~H.} \bibnamefont{Louisell}},
  \bibinfo{author}{\bibfnamefont{A.}~\bibnamefont{Yariv}}, \bibnamefont{and}
  \bibinfo{author}{\bibfnamefont{A.~E.} \bibnamefont{Siegman}},
  \bibinfo{journal}{Phys. Rev.} \textbf{\bibinfo{volume}{124}},
  \bibinfo{pages}{1646} (\bibinfo{year}{1961}).

\bibitem[{\citenamefont{Rubin et~al.}(1994)\citenamefont{Rubin, Klyshko, Shih,
  and Sergienko}}]{klyshko}
\bibinfo{author}{\bibfnamefont{M.~H.} \bibnamefont{Rubin}},
  \bibinfo{author}{\bibfnamefont{D.~N.} \bibnamefont{Klyshko}},
  \bibinfo{author}{\bibfnamefont{Y.~H.} \bibnamefont{Shih}}, \bibnamefont{and}
  \bibinfo{author}{\bibfnamefont{A.~V.} \bibnamefont{Sergienko}},
  \bibinfo{journal}{Phys. Rev. A} \textbf{\bibinfo{volume}{50}},
  \bibinfo{pages}{5122} (\bibinfo{year}{1994}).

\bibitem[{\citenamefont{Rubin}(1996)}]{Rubin1996}
\bibinfo{author}{\bibfnamefont{M.~H.} \bibnamefont{Rubin}},
  \bibinfo{journal}{Phys. Rev. A} \textbf{\bibinfo{volume}{54}},
  \bibinfo{pages}{5349} (\bibinfo{year}{1996}).

\bibitem[{\citenamefont{Law et~al.}(2000)\citenamefont{Law, Walmsley, and
  Eberly}}]{law2000}
\bibinfo{author}{\bibfnamefont{C.~K.} \bibnamefont{Law}},
  \bibinfo{author}{\bibfnamefont{I.~A.} \bibnamefont{Walmsley}},
  \bibnamefont{and} \bibinfo{author}{\bibfnamefont{J.~H.}
  \bibnamefont{Eberly}}, \bibinfo{journal}{Phys. Rev. Lett.}
  \textbf{\bibinfo{volume}{84}}, \bibinfo{pages}{5304} (\bibinfo{year}{2000}).

\bibitem[{\citenamefont{Hong et~al.}(1987)\citenamefont{Hong, Ou, and
  Mandel}}]{Hong1987}
\bibinfo{author}{\bibfnamefont{C.~K.} \bibnamefont{Hong}},
  \bibinfo{author}{\bibfnamefont{Z.~Y.} \bibnamefont{Ou}}, \bibnamefont{and}
  \bibinfo{author}{\bibfnamefont{L.}~\bibnamefont{Mandel}},
  \bibinfo{journal}{Phys. Rev. Lett.} \textbf{\bibinfo{volume}{59}},
  \bibinfo{pages}{2044} (\bibinfo{year}{1987}).

\bibitem[{\citenamefont{Huang and Eberly}(1993)}]{Huang1993}
\bibinfo{author}{\bibfnamefont{H.}~\bibnamefont{Huang}} \bibnamefont{and}
  \bibinfo{author}{\bibfnamefont{J.}~\bibnamefont{Eberly}},
  \bibinfo{journal}{J.~Mod.~Opt.} \textbf{\bibinfo{volume}{40}},
  \bibinfo{pages}{915} (\bibinfo{year}{1993}).

\bibitem[{\citenamefont{Kolenderski and Wasilewski}()}]{Kolenderski2009a}
\bibinfo{author}{\bibfnamefont{P.}~\bibnamefont{Kolenderski}} \bibnamefont{and}
  \bibinfo{author}{\bibfnamefont{W.}~\bibnamefont{Wasilewski}},
  \bibinfo{note}{in preparation}.


\end{thebibliography}

\end{document}